\documentclass[10pt, compress]{article}
\usepackage[labelfont=bf]{caption}

\usepackage{amsmath,amsfonts,amssymb,amsthm,bm,bbm,bbold,booktabs,color,epsfig,graphicx,hyperref,url}
\hypersetup{colorlinks=true, citecolor = blue}
\usepackage{mathrsfs}
\usepackage{lscape}
\usepackage{dcolumn}
\usepackage{tabularx}
\usepackage[round]{natbib}
\makeatletter
\providecommand*\AtBeginEnvironment[1]{%
  \@ifundefined{#1}%
    {\@latex@error{Environment #1 undefined}\@ehc
     \@gobble}%
    {\@ifundefined{ABE@env@#1}%
       {\expandafter\let\csname ABE@env@#1\expandafter\endcsname
          \csname #1\endcsname
        \expandafter\let\csname ABE@hook@#1\endcsname\@empty
        \@namedef{#1}{\@nameuse{ABE@hook@#1}\@nameuse{ABE@env@#1}}}%
       {}%
     \expandafter\g@addto@macro\csname ABE@hook@#1\endcsname}}
\@onlypreamble\AtBeginEnvironment
\makeatother

\AtBeginEnvironment{tabular}{\small}
\AtBeginEnvironment{align*}{\small}
\AtBeginEnvironment{align}{\small}

\newcommand{\Prob}{\mathbb{P}}

\usepackage{geometry}
\theoremstyle{plain}

\theoremstyle{definition}

\usepackage{lscape}
\usepackage[nameinlink,capitalise]{cleveref}

\usepackage{dcolumn}
\usepackage{tabularx}

\newcommand{\pkg}[1]{{\normalfont\fontseries{b}\selectfont #1}}
\let\proglang=\textsf
\let\code=\texttt

\usepackage{mathtools}

\DeclarePairedDelimiter{\diagfences}{(}{)}
\newcommand{\diag}{\operatorname{diag}\diagfences}

\usepackage{Sweave}
\begin{document}

\begin{titlepage}

\title{Multivariate ordinal regression for multiple repeated measurements}

\author{Laura Vana-G\"ur\thanks{TU Wien, Institute of Statistics and Mathematical Methods in Economics. Corresponding author. Email address: \url{laura.vana.guer@tuwien.ac.at}}}

\date{\today}
\maketitle

\begin{abstract}
In this paper we propose a multivariate ordinal regression
model which allows the joint modeling of three-dimensional panel data containing
both repeated and multiple measurements for a collection of subjects.
This is achieved by a multivariate autoregressive structure on
the errors of the latent variables underlying the ordinal responses,
where we distinguish between the correlations at a single point in time and the
persistence over time. The error distribution is assumed to be normal or 
Student~$t$ distributed.
The estimation is performed using composite likelihood methods. We perform
several simulation exercises to investigate the quality of the estimates in different settings
as well as in comparison with a Bayesian approach. The simulation study 
confirms that the estimation procedure is able to recover the model
parameters well and is competitive in terms of computation time. 
We also introduce \proglang{R} package 
\pkg{mvordflex} and illustrate how this implementation can be used to estimate the
proposed model in a user-friendly, convenient way.
Finally, we illustrate the framework on a data set containing firm failure and credit
ratings information from the rating agencies S\&P and Moody's for US listed companies.
\\
\vspace{0in}\\
\noindent\textbf{Keywords:} Composite likelihood,
Multivariate autoregressive error,
Multivariate ordinal regression model,
Panel data
\vspace{0in}\\
\bigskip
\end{abstract}

\end{titlepage}

%
%

\section{Introduction}
The analysis of correlated ordinal outcomes is an important task in a wide range
of research fields. It is often the case that multiple ordinal outcomes are
observed repeatedly over a period of time for a collection of subjects.
The modeling of such three-dimensional data (possibly in a regression setting)
should therefore take into account possible dependencies in the cross-section,
i.e.,~given that the multiple outcomes are observed on the same subjects, as
well as over time (i.e.,~longitudinal).

In this paper we propose a multivariate ordinal regression model which can
capture dependence among both repeated and multiple measurements in an ordinal
model. We achieve this by imposing a multivariate
AR(1) correlation structure on the errors of the continuous process underlying
the discrete ordinal observations.  The multivariate AR(1) process accounts for
the correlations among the multiple ordinal responses at the same point in time
as well as for the persistence in each of the multiple responses over time,
while keeping the number of parameters to be  estimated for the error structure
low. The model proposed in this paper therefore extends the two-dimensional class of
multivariate regression models to accommodate for a more complex dependence structure.

The estimation of the model parameters is performed by
composite likelihood methods.
In a simulation study we examine the quality of the estimates of the proposed
model for different scenarios related to the distribution of the errors and
to the degree of correlation present in the data. The results of the simulation
study confirm that the composite likelihood methods are able to recover the
parameters of the model well. We also perform a simulation exercise where we 
compare the proposed framework with an implementation using Bayesian methods and
show that for the investigated setting the pairwise likelihood approach is a 
competitive alternative to Bayesian inference, while 
having a lower computational cost. 

We implement the proposed model for three-dimensional panel data in \pkg{mvordflex}
\citep{pkgmvordflex}, which is built as an extension to the
existing \proglang{R} package \pkg{mvord}.
As in package \pkg{mvord}, the model can be estimated by using a multivariate
probit and multivariate logit link. Moreover, the regression coefficients and the
threshold parameters of the ordinal regression are allowed to vary across time
points and responses, but if more parsimonious specifications are desired, 
they can be constrained to be equal along some or all
time-outcome dimensions. Having a ready-to-use implementation will hopefully make
the model class more accessible to users in a variety of application fields.

In the empirical application we employ a data set of US listed firms which have been
rated by either Standard and Poor's (S\&P) or Moody's. For these firms we record
the available S\&P and Moody's ratings together with an indicator containing information
on whether the company went into bankruptcy in the year following the rating
observations. We show how the proposed model can be applied to these data and
how model comparison with simpler specifications can be performed.

The composite likelihood approach for estimation in multivariate ordinal
regression-type models is an attractive choice, given that it requires the
computation of low dimensional integrals instead of the high-dimensional
integrals necessary for the evaluation of the likelihood function.
Composite likelihood methods has been employed for ordinal models with
two-dimensional responses either in the cross-section
\citep[e.g.,][]{Scott02,bhat2010comparison,Pagui2015,Hirk2021jcr}
or longitudinally \citep[see e.g.,][]{Varin09,Reusens2017,tuzcuoglu2019,HIRK2022224}.
A software implementation for the two-dimensional model class
is provided in the package \pkg{mvord} for \proglang{R}
\citep{pub:mvord:Hirk+Hornik+Vana:2020}.

The estimation of regression models with autoregressive errors has been an 
active field of research. Among the pioneer papers which use linear regression 
with autoregressive errors to model a time-series in the presence of covariates 
we mention \cite{cochrane1949application,anderson1954problem,durbin1960estimation,zellner1964bayesian, chib1993bayes}. 
More recent papers include \cite{alpuim2008efficiency, tuacc2018robust, tuacc2020parameter} 
where the autoregressive errors of order~$p$ follow normal, Student-$t$ and skew-symmetric distributions respectively.
For the case of subjects observed over a collection of time points several approaches have been proposed
such as mixed effects models \citep{wang2010ecm} or joint mean-covariance models \citep{guney2022robust}.
For the three-dimensional setting, where several continuous outcomes (or responses) 
are observed for a collection of subjects  over time, the literature is scarcer, 
especially for modeling ordinal data, but several approaches have been proposed for different applications.
\cite{chaubert2008multivariate} propose a dynamic
multivariate ordinal probit model,
which assumes a general correlation structure
for the cross-sectional responses and a general multivariate autoregressive model on
the time-varying regression coefficients, rather than on the error terms.
This approach is appropriate if one assumes that the longitudinal
correlations arise due to the autocorrelation in the regression coefficients
rather than due to unobserved covariates. Similarly,
\cite{bartolucci2009multivariate} proposed multivariate model for categorical data where a set of subject-specific
intercepts assumed to follow a first-order Markov chain.
Another model class which has been employed are mixed-effect models,
which account for dependence in the responses by introducing latent effects at 
different levels of hierarchy. Conditional on these effects, the responses are 
typically assumed to be independent \citep[see e.g.,][]{LI201925}.
\cite{liu2006mixed} propose a three-level mixed effects item response model for
ordinal data which  contains subject and subject-time random intercepts. 
\cite{Lin2021} extend the model in \cite{liu2006mixed} to also include random 
slopes to measure the the effect of a subject on the response and also its change
over time.
\cite{cagnone2009latent} propose  latent variable models containing item-specific 
random effects and a common factor where the relationships between the 
time-dependent latent variables are modeled using autoregressive processes. 
A similar approach has been proposed in \cite{pub:djmdr:Vana+Hornik:2021} for a credit risk application
similar to the one presented in this paper.
An application of a model for three-dimensional panel data
is presented in \cite{SchliepSchaferHawkey+2021+241+254},
who employ a model with random coefficients to identify the cumulative effects of training and recovery in athletes.
The estimation of the
models presented above is typically performed by maximum likelihood methods
(using EM-type algorithms)
or, more commonly, by Markov chain Monte Carlo methods in a Bayesian setting,
where priors must be specified. In either case,
computations can prove to be rather intensive.

The paper is structured as follows: Section~\ref{sec:model} introduces the model
and Section~\ref{sec:simulation} presents the results of the simulation study.
Section~\ref{sec:empirical_results} introduces the credit risk application by describing
the data employed and presenting the results of the estimated model.
The software implementation as an \proglang{R} package is described in Section~\ref{sec:software}.
Section~\ref{sec:concl} concludes the paper.

\section{The model}\label{sec:model}
We extend the approach of multivariate ordinal regression models in
\cite{pub:mvord:Hirk+Hornik+Vana:2020}, which
model ordinal responses for a collection of
subjects observed either longitudinally (assuming an AR(1) correlation
structure, see e.g.,~\cite{HIRK2022224}) or
cross-sectionally (e.g.,~assuming a general correlation
structure, see e.g.,~\cite{Hirk2021jcr}).
In this paper we combine the two modeling settings by imposing a specific
multivariate autoregressive structure of order one on the errors of the latent
variables underlying the observed ordinal outcomes. The proposed
error structure is able to account for both
cross-sectional and longitudinal dependence among the ordinal responses.

\subsection{General set-up}
Let $y_{i,t}^j$ denote an ordinal observation and $i \in \{1, \ldots, n\}$ denotes
the subject index, $t \in \{1,2,\ldots,T\}$ is a time index among all equidistant
$T$ time points,
and $j \in \{1,\ldots q\}$ is the outcome index out
of all $q$ available outcomes.
Here we assume the panel data does not contain
any missing values (all outcomes in all time points are observed for all subjects).
Note however, that the framework can accommodate for missing values, as will be
discussed in Section~\ref{sec:subsec:missing}.
We assume the ordinal observation $y_{i,t}^j$ to be a coarser version of a continuous
latent variable $\tilde y_{i,t}^j$ connected by a vector of suitable
threshold parameters. These threshold parameters  $\bm \theta^j_t$ can in the most general case
be assumed to be time- as well as outcome-varying:
\begin{align*}
y_{i,t}^j = r \Leftrightarrow \theta^j_{t,r-1} < \tilde y_{i,t}^j \leq \theta^j_{t, r}, \quad r \in \{1, \ldots, K_j\},
\end{align*}
where $r$ is one of the $K_j$ ordered categories of outcome~$j$. For each
outcome~$j$ and time point~$t$, we have the monotonicity restriction on the
threshold parameters $\bm \theta_t^j$: $-\infty \equiv \theta_{t,0}^j < \theta_{t,1}^j < \cdots < \theta_{t,K_j-1}^j \equiv \infty$.
Furthermore, we assume the following linear regression model for
$\tilde y_{i,t}^j$:
\begin{align*}
  \tilde{y}_{i,t}^j = (\bm{x}_{i,t}^j)^\top {\bm \beta}_{t}^j + \epsilon_{i,t}^j
\end{align*}
i.e.,~$\tilde y_{i,t}^j$ depends linearly on a $p$-dimensional vector of time- and
outcome-specific covariates $\bm{x}_{i,t}^j$, where ${\bm \beta}_{t}^j$ is a 
time- and outcome-specific $p$-dimensional vector of covariates 
and $\epsilon_{i,t}^j$ is an error term of subject $i$ for outcome $j$ in time $t$.

In the complete case $\bm y_{i,t}$ is a $q$-dimensional vector with
$\bm y_{i,t} = (y_{i,t}^1, y_{i,t}^2, \ldots, y_{i,t}^{q})^\top$. We define the following
$(p\cdot q)\times q$ matrix of predictors:
\begin{align*}
X_{i,t}^* =
\begin{pmatrix} (\bm x_{i,t}^1)^\top & \bm 0 & \cdots & \bm 0\\
  \bm 0 & (\bm x^2_{i,t})^\top & \cdots & \bm 0\\
  \vdots & \vdots & \ddots & \vdots\\
  \bm 0 & \bm 0 & \cdots & (\bm x^q_{i,t})^\top\\
\end{pmatrix}.
\end{align*}
Assuming $\bm \beta_t^*$ to be a $p\cdot q$-dimensional vector
$\bm \beta_t^* =((\bm{\beta}_{t}^1)^\top, (\bm{\beta}_{t}^2)^\top, \ldots, (\bm{\beta}_{t}^q)^\top)^\top$,
the $q$-dimensional latent process for each $i$ and $t$ is given by:
\begin{align*}
{\bm {\tilde y}}_{i,t} =  X_{i,t}^* \bm \beta_t^* + \bm \epsilon_{i,t},
\end{align*}
where $\bm \epsilon_{i,t}=(\epsilon^1_{i,t},\ldots,\epsilon^q_{i,t})^\top$ is a
$q$-dimensional vector of errors.

\subsection{Structure of the errors}\label{sec:model_error}
We consider an auto-regressive structure on the $q$-dimensional error terms $\bm \epsilon_{i,t}$:
\begin{align}\label{eq:ar1_errors}
 \bm  \epsilon_{i,t} = \Psi \bm \epsilon_{i, t-1} + \Sigma^{1/2}_t \bm u_{i,t},
\end{align}
where $\Psi =  \diag{\psi_1, \psi_2, \ldots, \psi_{q}}$ is a diagonal matrix
of persistence parameters for each outcome $j$ with $\vert \psi_j\vert<1$.
These persistence parameters will capture the
longitudinal dependence on past values of the same outcome\footnote{Alternatively,
if the errors should also depend on lags of other outcomes,
$\Psi$ can be chosen to be a full matrix with eigenvalues smaller than one in
absolute value to ensure stationarity.
We leave the implementation and investigation of such a specification for future research.}.
The $q$-dimensional mean-zero error term $\bm u_{i,t}$ is independent and identically
distributed among the
subjects, outcomes and time points with cumulative distribution function 
$F$ ($u^j_{i,t}\stackrel{iid}\sim F$) and it is independent of $\epsilon^j_{i,t-k}, k>0$ for all $t$ and $j$.
The matrix $\Sigma_t$ captures the
cross-sectional correlation among the different outcomes at time $t$ conditional on $\bm\epsilon_{i,t-1}$.

Assume that $Y_i$ is a $q \times T$ matrix and let $\bm y_i^*$ be the vectorization of the matrix $Y_i$:
\begin{align*}
 Y_i =  (\bm y_{i,1}, \bm y_{i,2}, \ldots, \bm y_{i,T}) = \begin{pmatrix}
y_{i,1}^1 & y_{i,2}^1 & \cdots &  y_{i,T}^1\\
y_{i,1}^2 & y_{i,2}^2 & \cdots &  y_{i,T}^2\\
\vdots & \vdots & \ddots & \vdots\\
y_{i,1}^q & y_{i,2}^q & \cdots &  y_{i,T}^q\\
\end{pmatrix},
\quad  \bm y_i^* = \text{vec}(Y_i) =  (y_{i,1}^1, \ldots,  y_{i,1}^{q}, y_{i,2}^1, \ldots,
              y_{i,2}^{q},  \ldots, y_{i,T}^1, \ldots,  y_{i,T}^{q})^\top.
\end{align*}
For the corresponding vector of latent variables $\bm {\tilde y}_{i}^*$ we have:
\begin{align}\label{eqn:mvord}
  \bm {\tilde y}_{i}^* =  X_{i}^{*} \bm \beta^{*} + \bm \epsilon_{i}^*,
\end{align}
where $X_{i}^{*}$ is a block-diagonal matrix with
\begin{align*}
X_{i}^{*} =
\begin{pmatrix}
X_{i,1}^* & \bm 0 & \cdots & \bm 0\\
\bm 0 &  X_{i,2}^* & \cdots & \bm 0\\
\vdots & \vdots & \ddots & \vdots\\
\bm 0 & \bm 0 & \cdots & X_{i,T}^*\\
\end{pmatrix},
\end{align*}
$\bm \beta^{*}$ is a $pqT$-dimensional vector of regression coefficients
$\bm \beta^{*} =((\bm{\beta}_{1}^*)^\top, (\bm{\beta}_{2}^*)^\top, \ldots, (\bm{\beta}_{T}^*)^\top)^\top$ and 
 $\bm\epsilon_i^*=(\bm \epsilon_{i,1}^\top, \bm  \epsilon_{i,2}^\top, \ldots, \bm \epsilon_{i,T}^\top)^\top$ denote the subject-level $qT$-dimensional mean zero error terms.
 
We are interested in representing the above model at a
subject level, where the dependence in the subject-level errors 
is given by the stationary distribution of the process in
Equation~\eqref{eq:ar1_errors}.

\subsubsection{Multivariate probit link}
A common approach is to assume that $u^j_{i,t}$ has a standard normal
distribution, and that the conditional covariance matrix $\Sigma_t$ is constant. Here we assume
\begin{align*}
\Sigma_t=
 \Sigma = \begin{pmatrix} 1 & \rho_{1,2} & \cdots & \rho_{1,q}\\
  \rho_{1,2}& \ddots & \ddots & \rho_{2,q}\\
  \vdots & \ddots & \ddots & \vdots\\
  \rho_{1,q}&  \rho_{2,q}& \cdots & 1
  \end{pmatrix}.
\end{align*}
 has a general correlation structure with $q(q-1)/2$ parameters to be estimated, with  diagonal elements
 set to one to ensure identifiability in the marginal
ordinal models. 

Then, $\bm\epsilon^*_i$ follows a multivariate normal distribution with mean zero and covariance matrix
\begin{align}\label{eq:corstruct}
\bm\epsilon^*_i=MVN_{qT}(\bm 0, \Sigma^*), \quad \Sigma^*=
\begin{pmatrix}
 \tilde\Sigma  & (\Psi\tilde\Sigma)^\top &  (\Psi^2  \tilde\Sigma)^\top & \cdots  & (\Psi^{T-1}  \tilde\Sigma)^\top\\
\Psi  \tilde\Sigma & \tilde\Sigma  &  (\Psi\tilde\Sigma)^\top  & \cdots &  (\Psi^{T-2}\tilde\Sigma)^\top\\
\Psi^2 \tilde\Sigma & \Psi \tilde\Sigma &\tilde\Sigma & \cdots &  (\Psi^{T-3} \tilde\Sigma)^\top\\
\vdots & \ddots &  \ddots & \ddots & \vdots\\
\Psi^{T-1} \tilde\Sigma & \cdots & \cdots & \Psi\tilde\Sigma & \tilde\Sigma
\end{pmatrix},
\end{align}
where $\tilde\Sigma$  the unconditional variance of the multivariate AR(1) process in 
Equation~\eqref{eq:ar1_errors} given by $vec(\tilde\Sigma)=(I-\Psi\otimes\Psi)^{-1}vec(\Sigma)$.
This choice of $F$ gives rise to the multivariate probit link.
\subsubsection{Multivariate logit link}\label{sec:mvlink_logit}
In this section we show how we choose $F$ in a way that gives rise to the multivariate logit link 
in \cite{pub:mvord:Hirk+Hornik+Vana:2020}, where the subject-level errors are
assumed to have the multivariate logistic distribution of \cite{OBrien2004}
$\mathcal{L}_{qT}(0, \Sigma^*)$. 
We use the fact that the multivariate Student $t$ distribution closely 
approximates the multivariate logistic distribution in \cite{OBrien2004} when
the covariance matrix $\Sigma^*$ is scaled by a constant $\gamma=\pi^2(\nu-2)/(3\nu)$ and $\nu \approx 8$. 
Therefore, we choose $F$ such that $\bm\epsilon^*_i$ follows a $qT$-variate 
Student-$t$ distribution with $\nu$ degrees of freedom, mean zero and covariance matrix $\gamma\Sigma^*$:
\begin{align*}
\bm\epsilon^*_i \sim MVT_{qT}(\bm 0, \gamma\Sigma^*, \nu), \qquad \nu=8.
\end{align*}
In order to achieve this stationary distribution on $\bm\epsilon^*_i$, we can apply 
the results in \cite{virolainen2021gaussian} and choose $F$ as the univariate 
Student $t$ distribution with scale one and $\nu + q$ degrees of freedom.
Unlike in the normal distribution case, the conditional covariance
$\Sigma_t$ is heteroscedastic in the multivariate AR(1) process with Student $t$ errors:
\begin{align*}
u^j_{i,t} \sim t(0, 1, \nu + q), \quad
\bm\epsilon_{i,0} \sim MVT_q(\bm 0, \gamma\tilde\Sigma, \nu),\quad
\Sigma_t &= \frac{\nu-2 + \bm\epsilon_{i,t-1}^\top(\gamma\tilde\Sigma)^{-1}\bm\epsilon_{i,t-1}}{\nu-2+q}\gamma\Sigma.
\end{align*}
According to Theorem 1 of \cite{virolainen2021gaussian}, this implies a multivariate Student~$t$ stationary distribution on the errors
$\bm\epsilon_{i,t} \sim MVT_q(\bm 0, \gamma\tilde\Sigma, \nu)$. This in turn translates into $\bm\epsilon^*_i\sim MVT_{qT}(\bm 0, \gamma\tilde\Sigma^*, \nu)$. Note that in all exercises involving the approximate logit link we fix $\nu=8$.
%
%
%
%

\subsection{Pairwise likelihood estimation and inference}\label{sec:inference}
For a given vector of parameters $\bm\delta$ containing
the threshold parameters, regression coefficients and parameters of the error structure,
the likelihood is given by the product of the following multivariate probabilities
over all subjects:
\begin{align}\label{eqn:likfct}
L(\bm\delta; Y, X) &= \prod_{i=1}^n \Prob \bigg(\bigcap_{\substack{j\in
   {1,\ldots,q}\\  t \in \{1, \ldots, T\}}}
   \{y_{i,t}^j = r_{i,t}^j\}\vert X_i\bigg).
\end{align}
In the complete case, each multivariate probability corresponds to a
$q\times T$-dimensional integral
\begin{align*}
\Prob \bigg(\bigcap_{\substack{j\in
   {1,\ldots,q}\\  t \in \{1, \ldots, T\}}} \{y_{i,t}^j = r_{i,t}^j\}\vert X\bigg) =
 \int_{D_{i}} f_{qT}(\bm{\tilde{y}^*}_{i}; \bm \delta, X^*_i) d^{qT}
 \bm{\tilde{y}^*}_{i},
\end{align*}
where $D_{i} =
\prod_{t\in \{1, \ldots, T\}} \prod_{j\in{1,\ldots,q}}  (\theta^j_{t,r_{i,t}^j-1},
\theta^j_{t,r_{i,t}^j})$ is a
  Cartesian product (here $r_{i,t}^j$ denotes the observed ordinal class for subject~$i$,
  time~$t$ and outcome~$j$)
  and $f_{qT}$
  is the multivariate density of the error terms.

For parameter estimation of model \eqref{eqn:mvord} we use a composite
likelihood approach, where we approximate the full likelihood above by a pairwise likelihood which is
constructed from bivariate marginal distributions. The pairwise
likelihood function is given by the product of the bivariate probabilities
corresponding to all pairs of elements in $\bm y_i^*$:
\begin{align}\label{eqn:logpl}
  PL(\bm\delta; Y, X)= \prod_{i=1}^n\prod_{k=1}^{(q\cdot T)-1}
  \prod_{l=k+1}^{q\cdot T}PL_i^{(k,l)}(\bm\delta;Y,X), \quad PL_i^{(k,l)}(\bm\delta;Y_i,X_i)= \Prob\biggl((\bm y_{i}^*)_k = (\bm r_{i})_k,
    (\bm y_{i}^*)_l = (\bm r_{i})_l\vert X^*_i\biggr),
\end{align}
where $(\bm y_{i}^*)_k$ denotes the $k$-th element of vector $\bm y^*_{i}$ and
 $(\bm r_{i})_k$ denotes the $k$-th element of subject-specific vector
 $\bm r_{i}= (r_{i,1}^1, \ldots,  r_{i,1}^{q}, r_{i,2}^1, \ldots, r_{i,2}^{q},
 \ldots, r_{i,T}^1, \ldots,  r_{i,T}^{q})^\top$.

Given that the pairwise likelihood for subject~$i$ consists of the product of
${q\cdot T\choose 2}$ bivariate probabilities, for cases where $q\cdot T$ is large
this can prove to be computationally burdensome. In order to speed-up the pairwise
likelihood computation, one option is to only consider pairs which lie close
to each other in time, as they are the ones who are most informative on the
persistence parameter. With a slight abuse of notation we denote $k_t$
the time index and $k_j$ the outcome index corresponding to the $k$-th element 
in the vector $(\bm y_{i}^*)$ (e.g.,~for $k=2$ $k_t=1$ and $k_j = 2$). The expression
in \eqref{eqn:logpl} is adjusted to:
\begin{align}\label{eqn:logpladj}
  PL_i^{(k,l)}(\bm\delta, c;Y_i,X_i), =\biggl[\Prob\biggl((\bm y_{i}^*)_k = (\bm r_{i})_k,
    (\bm y_{i}^*)_l = (\bm r_{i})_l\vert X_i\biggr) \biggr]^{\mathbb 1(l_t - k_t \leq c)}
\end{align}
where $c$ is a pre-defined lag and $\mathbb 1$ is the indicator function.  This
strategy of considering only pairs of observations less distant than $c$ time
points has also been employed in \cite{Varin09}, who
propose tuning this parameter as the value minimizing a global
fitting criterion such as the generalized variance (the determinant of
the estimated covariance matrix of the estimates).

The maximum pairwise likelihood estimates $\hat{\bm\delta}_{PL}(c)$ are obtained by direct maximization of the log pairwise likelihood
using general purpose optimization tools. 
Under regularity conditions, $\sqrt{n}(\hat{\bm\delta}_{PL}(c)-\bm\delta)$
has an asymptotic normal distribution with mean 0 and covariance matrix
equal to the Godambe information matrix $G(\bm\delta, c)=H(\bm\delta,c)^{-1}V(\bm\delta,c)H(\bm\delta,c)^{-1}$ for a fixed value of $c$. The following consistent estimates of the Hessian matrix $H(\bm\delta,c)$ and variability matrix $V(\bm\delta,c)$ only necessitate the first derivatives of the log likelihood with respect to the parameters.
\begin{align*}
\hat H(\bm\delta,c)=\sum_{i=1}^n\sum_{k=1}^{(q\cdot T)-1}
  \sum_{l=k+1}^{q\cdot T}\left(\frac{\partial \log PL_{i}^{(k,l)}(\bm\delta, c;Y_i,X_i)}{\partial\bm\delta}\right)\left(\frac{\partial \log PL_{i}^{(k,l)}(\bm\delta, c;Y_i,X_i)}{\partial\bm\delta}\right)^\top
\end{align*}
\begin{align}\label{eq:variability_matrix}
\hat V(\bm\delta,c)=\sum_{i=1}^n\left(\sum_{k=1}^{(q\cdot T)-1}
  \sum_{l=k+1}^{q\cdot T}\frac{\partial \log PL_{i}^{(k,l)}(\bm\delta, c;Y_i,X_i)}{\partial\bm\delta}\right)\left(\sum_{k=1}^{(q\cdot T)-1}
  \sum_{l=k+1}^{q\cdot T}\frac{\partial \log PL_{i}^{(k,l)}(\bm\delta, c;Y_i,X_i)}{\partial\bm\delta}\right)^\top.
\end{align}
In general, adding more bivariate likelihood components will increase the 
Hessian matrix and therefore reduce the variance but adding too many correlated 
bivariate likelihoods will inflate the variability matrix \citep{ferrari2016parsimonious}. 
Moreover, in finite samples, the estimator $\hat H(\bm\delta,c)$ but more so 
$\hat V(\bm\delta,c)$ is unstable when $n$ is rather small compared to the 
number of parameters \citep{varin_overview}. Alternatives to the estimator in
Equation~\eqref{eq:variability_matrix} include re-sampling methods such as 
bootstrap, jackknife, or the less computationally demanding one-step jackknife 
which offers a first order approximation to the jackknife 
\citep[see][]{varin_overview, ferrari2016parsimonious}.

Finally, model comparison can be performed using information criteria such as
the composite likelihood Akaike or Bayesian information criterion
\citep[for more details see e.g.,][]{Varin2005}.

\subsection{Missing values}\label{sec:subsec:missing}
In the presence of missing values in the $q\times T$ vector of responses of subject~$i$,
we employ the same strategy as \cite{pub:Hirk+Hornik+Vana:2018a} and
construct the pairwise likelihood only from the bivariate probabilities
corresponding to all pairs of \emph{observed} responses. If the number of
observed outcomes for subject~$i$ is less than two, the univariate marginal
distribution enters the likelihood instead of the bivariate ones.
This approach assumes that the missing value mechanism is completely at random.
Approaches to model the missing data mechanism jointly with the 
observations in longitudinal models can be found in e.g.,~\cite{li2013pairwise, LI201925}.

\subsection{Constraints on the threshold and regression coefficients}\label{sec:subsec:constraints}

Constraints can be set on the vector of regression coefficients and on
the threshold parameters,
which in the most general case are assumed to be time- and outcome-varying.
We define two $q\times T$-dimensional linear predictors by making again use of the matrix notation:
\begin{align*}
\bm\eta^\text{upper}_{i}= B_i^\text{upper}\bm\theta^* - X_i^*\bm \beta^* = Z^{\text{upper}}_i\bm\kappa^* ,
\quad
\bm\eta^\text{lower}_{i}= B_i^\text{lower}\bm\theta^* - X_i^*\bm \beta^*= Z^{\text{lower}}_i\bm\kappa^*,
\quad
Z^{\text{.}}_i=(B_i^\text{.}, -X_i^*),
\quad
\bm\kappa^*=((\bm\theta^*)^\top, (\bm\beta^*)^\top)^\top,
\end{align*}
where
$\bm\theta^*=((\bm{\theta}^1_{1})^\top,\ldots,(\bm{\theta}^q_{1})^\top, \ldots,
(\bm{\theta}^1_{T})^\top,\ldots,(\bm{\theta}^q_{T})^\top)^\top
$ and the matrices $B_i^\text{lower}$ and $B_i^\text{upper}$ are
$(q\times T)\times (T\sum_{j=1}^q (K_j-1))$ block diagonal binary matrices
\begin{align*}
B^\text{upper}_{i}&=\diag{
  (\bm b_{i,1}^{1,\text{upper}})^\top,\ldots,
  (\bm b_{i,1}^{q,\text{upper}})^\top, \ldots,
  (\bm b_{i,T}^{1,\text{upper}})^\top,\ldots,
  (\bm b_{i,T}^{q,\text{upper}})^\top}\\
  B^\text{lower}_{i}&=\diag{
  (\bm b_{i,1}^{1,\text{lower}})^\top,\ldots,
  (\bm b_{i,1}^{q,\text{lower}})^\top, \ldots,
  (\bm b_{i,T}^{1,\text{lower}})^\top,\ldots,
  (\bm b_{i,T}^{q,\text{lower}})^\top}
\end{align*}
where the vector $\bm b_{i,t}^{j,\text{upper}}$ has length $K_j-1$ and contains a one in the $r_{i,t}^j$-th position if $r_{i,t}^j\in \{1,\ldots,K_j-1\}$, else zero; the vector $\bm b_{i,t}^{j,\text{lower}}$ has length $K_j-1$ and contains a one in the $(r_{i,t}^j - 1)$-th position if $r_{i,t}^j\in \{2,\ldots,K_j\}$, else zero.

The probabilities in the likelihood function in Equation~\eqref{eqn:likfct} can then be expressed as:
\begin{align*}
\Prob \bigg(\bigcap_{\substack{j\in
   {1,\ldots,q}\\  t \in \{1, \ldots, T\}}} \{y_{i,t}^j = r_{i,t}^j\}\bigg) = F_{qT}(Z_i^\text{upper}\bm\kappa^*|\Sigma^*, \ldots) - F_{q T}(Z_i^\text{lower}\bm\kappa^*|\Sigma^*, \ldots).
\end{align*}

Assuming that $\tilde{\bm\kappa} = (\tilde{\bm\theta}^\top, \tilde{\bm\beta}^\top)^\top$ is the reduced $(h\times 1)$ vector of thresholds and coefficients to be estimated, the linear predictors can be rewritten as:
\begin{align*}
\bm\eta^\text{.}_{i}=  Z^{\text{.}}_i C \tilde{\bm\kappa}
\end{align*}
where $C$ is a contrast matrix of dimension $(T\times\sum_{j=1}^q(K_j-1) + q T p) \times h$. For example, the $C$ matrix for a model where all thresholds should be constant over time and one set of regression coefficients should be employed for all $t$ and $j$ would
be of dimension
$(T\sum_{j=1}^q(K_j-1) + q T p) \times (\sum_{j=1}^q(K_j-1) + p)$:
\begin{align*}
C=\left(\begin{array}{c c}
\underbrace{(1,\ldots,1)^\top }_{T\, \text{times}}\otimes
I_{\sum_{j=1}^q(K_j-1)}
& \bm 0_{T\sum_{j=1}^q(K_j-1) \times p}\\
\bm 0_{ (q \cdot T\cdot p) \times p}&
\underbrace{(1,\ldots,1)^\top }_{q\cdot T\, \text{times}}\otimes  I_{p}
\end{array}
\right),
\end{align*}
where $I_{.}$ denotes the identity matrix, $\bm 0$ is the zero matrix and
$\otimes$ denotes the Kronecker product.
%
\section{Simulation study}\label{sec:simulation}

In order to investigate the quality of pairwise likelihood estimates of the
proposed model we perform a simulation study. Within this study we simulate
data from the proposed model with various parameter settings.

\subsection{Different correlation settings}\label{sec:sim_big}
In a first exercise, we investigate the performance of the pairwise likelihood 
estimates when we employ different parameter values for the correlation
structure. We consider a moderate data setting with $n=1000$ subjects. 
To align with the setting of the application presented in 
Section~\ref{sec:empirical_results}, we generate a panel data with $q=3$
different outcomes and $T=10$ time points.
Two of the ordinal outcomes have four categories and one outcome is binary
($K_1 = 4$, $K_2 = 4$, $K_3=2$).
The threshold parameters vary among the three outcomes but are assumed to be 
constant over all time points:
$\bm\theta^1=(-\infty,- 3, 0, 3,\infty)$,
$\bm\theta^2=(-\infty,- 2 , 0, 2,\infty)$,
$\bm\theta^3=(-\infty, 3, \infty)$.
For the first outcome, we choose the thresholds such that the distribution 
of the four categories is more concentrated in the middle categories $2$ and $3$
and less on the peripheral categories $1$ and $4$. 
For the second outcome, the chosen thresholds lead to a more balanced distribution.
The binary outcome is imbalanced.
In all settings, we simulate $p = 2$ covariates from a standard normal 
distribution, which vary for each of the $T$ time points, 
but do not vary with the outcomes $j=1,2,3$. Again, this is in line with the
empirical application where the covariates are built from the 
financial information of corporations and do not vary with the 
creditworthiness indicators used as response variables.
The vector of regression coefficients is constant among all time points and outcomes
$\bm\beta=(2,-1)^\top$.
For the error structure 
we simulate four different combinations of the inter-rater correlation matrix
$\Sigma$ and the time-persistence matrix $\Psi$:
\begin{align*}
  \Sigma_\text{low} =
  \begin{pmatrix}
    1.000 & 0.100 & 0.200\\
    0.100 & 1.000 & 0.300\\
    0.200 & 0.300 & 1.000
  \end{pmatrix},&\qquad
 \Sigma_\text{high} =
  \begin{pmatrix}
    1.000 & 0.950 & 0.875\\
    0.950 & 1.000 & 0.800\\
    0.875 & 0.800 & 1.000
  \end{pmatrix}\\
  \Psi_\text{low} =
    \begin{pmatrix}
    0.200 & 0& 0\\
    0 & 0.250 & 0\\
    0 & 0 & 0.350
  \end{pmatrix},&\qquad
  \Psi_\text{high} =
  \begin{pmatrix}
    0.800 & 0 & 0\\
    0 & 0.850 & 0\\
    0 & 0 & 0.900
  \end{pmatrix}
\end{align*}
In previous simulation studies on the performance of the pairwise likelihood 
estimates in ordinal regression models it has been observed that, 
when the true correlation
among responses is low, these correlation parameters are not recovered 
as well as when the correlation is high 
\citep[see study in e.g.,][]{pub:Hirk+Hornik+Vana:2018a}. 
To investigate whether this is also the case in the proposed model,
we consider four different scenarios for the error structure
$\Sigma$ low -- $\Psi$ low, $\Sigma$ low -- $\Psi$ high,
$\Sigma$ high -- $\Psi$ low, $\Sigma$ high -- $\Psi$ high.
Finally, we perform the simulation study for the multivariate probit link
 multivariate logit link introduced in Section~\ref{sec:model_error}.
In total we therefore consider 8 scenarios. In all scenarios we assume $c=T-1$
do not exclude any pairs of observations from the pairwise likelihood.
\begin{figure}
\centering
\includegraphics{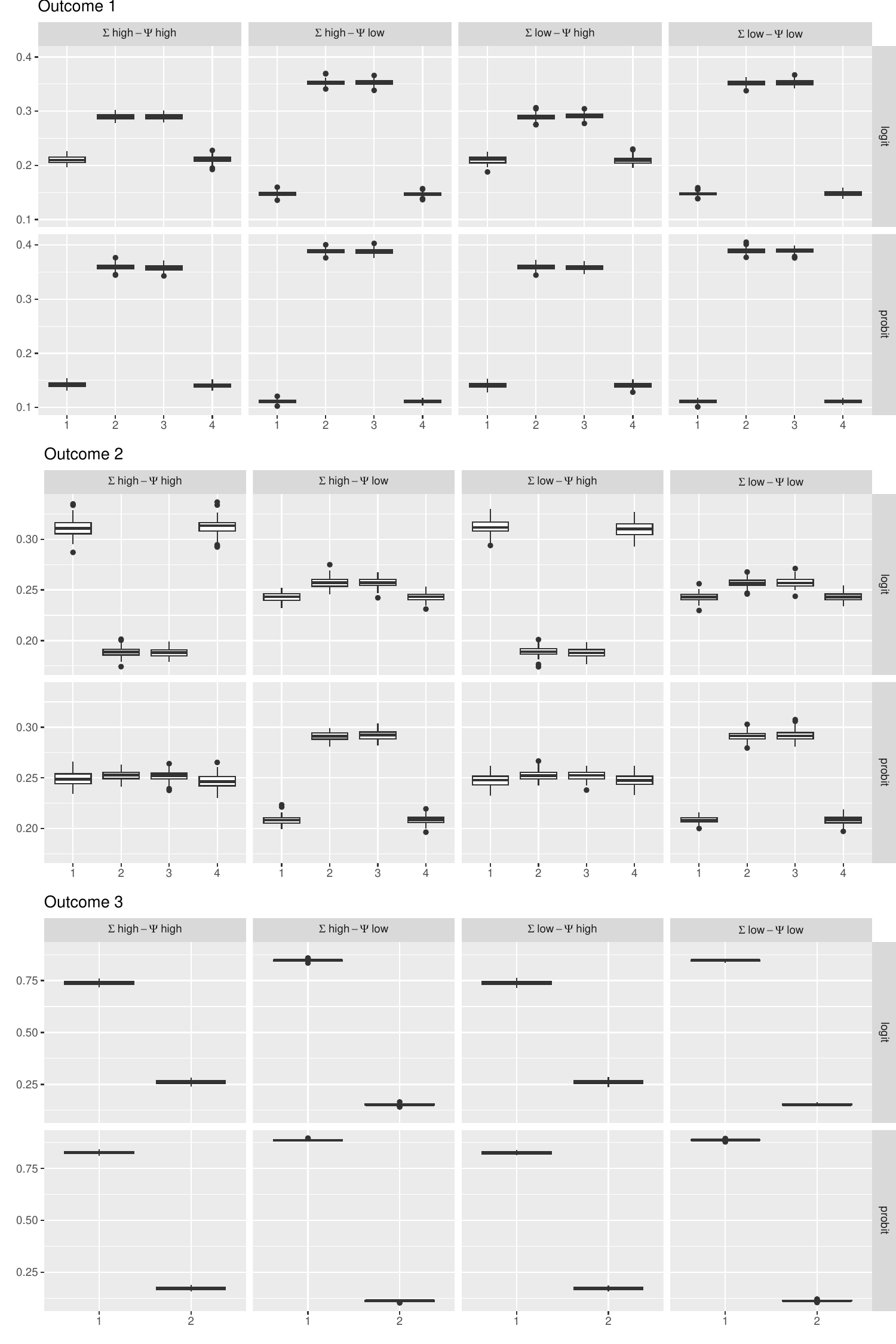}
\caption{This figure displays the distribution of the categories for each 
outcome over the 100 simulated data sets for the 8 scenarios considered in the 
simulation exercise with $n=1000$ subjects, $q=3$ responses 
(two ordinal responses with four classes and a binary response) observed over 
$T=10$ time points and $p=2$ covariates.}
\label{fig:sim_stats_response}
\end{figure}
For all scenarios we replicated the simulation of the data sets 100 times.
In Figure~\ref{fig:sim_stats_response} we present for each outcome 
the distribution of the categories over the 100 data sets for each of the
8 scenarios. 

\begin{figure}
\centering
\includegraphics{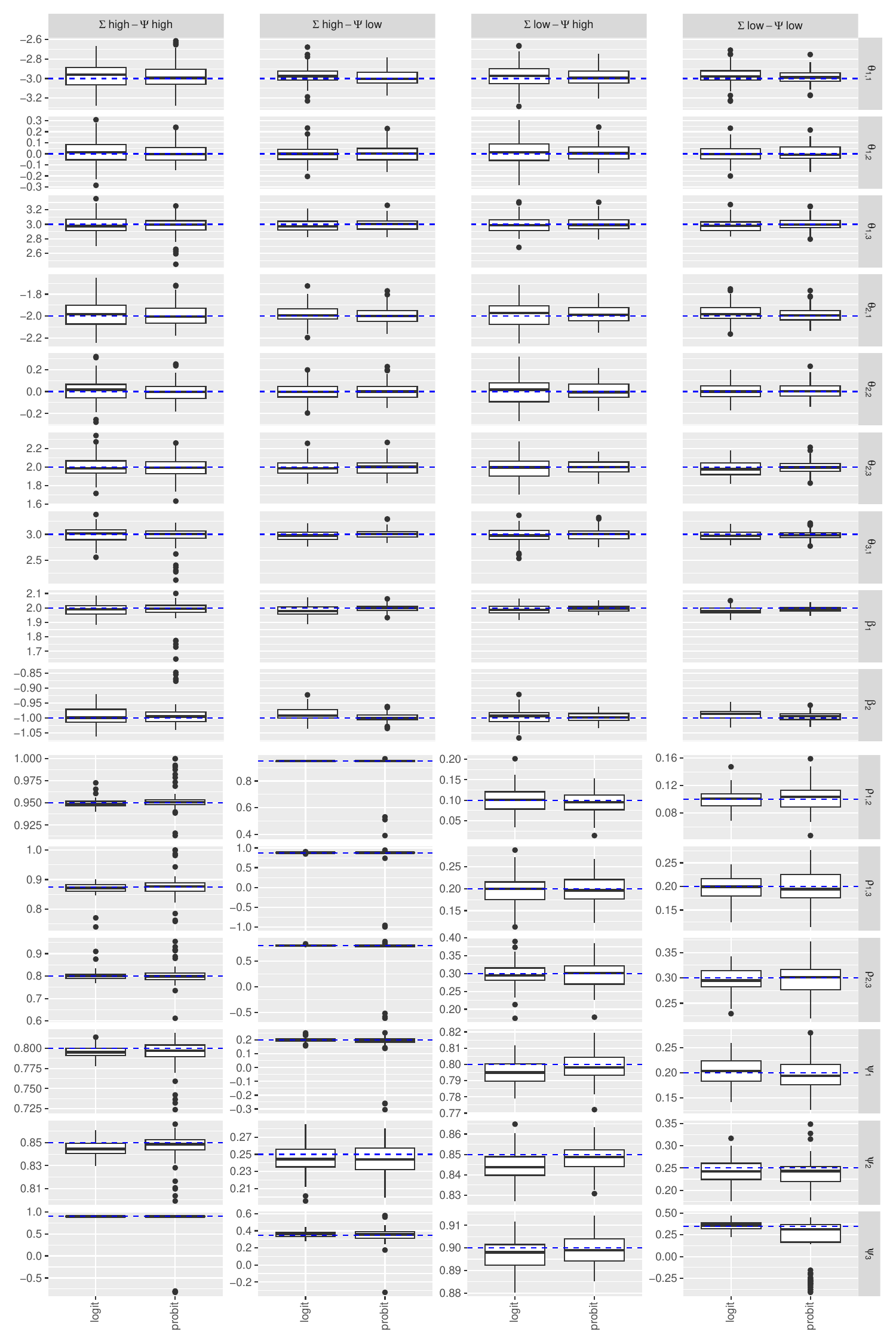}
\caption{This figure displays the results of the simulation exercise with 
$n=1000$ subjects, $q=3$ responses (two ordinal responses with four classes 
and a binary response) observed over  $T=10$ time points and $p=2$ covariates.
The estimates over the 100 simulated data sets for the 4 correlation scenarios 
and two link functions are illustrated for all model parameters. }
\label{fig:sim_big_all}
\end{figure}

Estimates of the parameters for the four correlation scenarios and two link functions
are illustrated in Figure~\ref{fig:sim_big_all}. 
Furthermore, we calculated the mean parameter estimate of the repetitions,
the absolute percentage bias (APB)
\footnote{$\text{APB} = |(\text{true parameter} - \text{mean estimate})/\text{true parameter}|$. 
Note that for values of zero we do not report the APB.},
the mean asymptotic standard error and the standard deviation of the parameters
over the 100 repetitions. These results are presented in 
\cref{tab:sim_t_10_logit_probit_id_4,tab:sim_t_10_logit_probit_id_3,tab:sim_t_10_logit_probit_id_2,tab:sim_t_10_logit_probit_id_1}.
\begin{table}[ht]
\centering
\caption{This table presents simulation results based on 100 repetitions, $n=1000$, $T = 10$, $q=3$ for the correlation setting $\Sigma$ high and $\Psi$ high.} 
\label{tab:sim_t_10_logit_probit_id_1}
\begin{tabular}{rrrlrrrlrr}
  \toprule
  & & \multicolumn{4}{c}{Multivariate probit link}& \multicolumn{4}{c}{Multivariate logit link}\\
                             \cmidrule(lr){3-6}\cmidrule(lr){7-10}\\& True &Mean Est&APB&\multicolumn{1}{p{1.4cm}}{\centering Mean Asym SE}&SD Sample&Mean Est&APB&\multicolumn{1}{p{1.4cm}}{\centering Mean Asym SE}&SD Sample\\ \midrule
$\theta_{1,1}$ & $-$3.0000 & $-$2.9767 & 0.78\% & 0.0632 & 0.1076 & $-$2.9719 & 0.94\% & 0.0771 & 0.1278 \\ 
  $\theta_{1,2}$ & 0.0000 & 0.0005 & \multicolumn{1}{c}{\quad -} & 0.0430 & 0.0734 & 0.0130 & \multicolumn{1}{c}{\quad -} & 0.0425 & 0.1098 \\ 
  $\theta_{1,3}$ & 3.0000 & 2.9792 & 0.69\% & 0.0624 & 0.1032 & 2.9956 & 0.15\% & 0.0776 & 0.1164 \\ 
  $\theta_{2,1}$ & $-$2.0000 & $-$1.9907 & 0.46\% & 0.0597 & 0.0949 & $-$1.9807 & 0.96\% & 0.0666 & 0.1219 \\ 
  $\theta_{2,2}$ & 0.0000 & $-$0.0025 & \multicolumn{1}{c}{\quad -} & 0.0516 & 0.0738 & 0.0096 & \multicolumn{1}{c}{\quad -} & 0.0501 & 0.1127 \\ 
  $\theta_{2,3}$ & 2.0000 & 1.9867 & 0.66\% & 0.0595 & 0.0934 & 1.9998 & 0.01\% & 0.0672 & 0.1091 \\ 
  $\theta_{3,1}$ & 3.0000 & 2.9719 & 0.94\% & 0.0846 & 0.1379 & 2.9979 & 0.07\% & 0.0959 & 0.1520 \\ 
  $\beta_{1}$ & 2.0000 & 1.9866 & 0.67\% & 0.0338 & 0.0437 & 1.9878 & 0.61\% & 0.0442 & 0.0431 \\ 
  $\beta_{2}$ & $-$1.0000 & $-$0.9932 & 0.68\% & 0.0219 & 0.0254 & $-$0.9938 & 0.62\% & 0.0265 & 0.0304 \\ 
  $\rho_{1,2}$ & 0.9500 & 0.9523 & 0.24\% & 0.0044 & 0.0086 & 0.9495 & 0.05\% & 0.0055 & 0.0048 \\ 
  $\rho_{1,3}$ & 0.8750 & 0.8758 & 0.09\% & 0.0178 & 0.0229 & 0.8705 & 0.52\% & 0.0215 & 0.0217 \\ 
  $\rho_{2,3}$ & 0.8000 & 0.8027 & 0.34\% & 0.0195 & 0.0275 & 0.8003 & 0.04\% & 0.0228 & 0.0190 \\ 
  $\psi_{1}$ & 0.8000 & 0.7955 & 0.57\% & 0.0087 & 0.0111 & 0.7955 & 0.57\% & 0.0114 & 0.0078 \\ 
  $\psi_{2}$ & 0.8500 & 0.8470 & 0.36\% & 0.0069 & 0.0081 & 0.8451 & 0.58\% & 0.0088 & 0.0071 \\ 
  $\psi_{3}$ & 0.9000 & 0.8454 & 6.07\% & 0.0065 & 0.2467 & 0.8975 & 0.28\% & 0.0087 & 0.0063 \\ 
   \bottomrule
\end{tabular}
\end{table}
\begin{table}[ht]
\centering
\caption{This table presents simulation results based on 100 repetitions, $n=1000$, $T = 10$, $q=3$ for the correlation setting $\Sigma$ high and $\Psi$ low.} 
\label{tab:sim_t_10_logit_probit_id_2}
\begin{tabular}{rrrlrrrlrr}
  \toprule
  & & \multicolumn{4}{c}{Multivariate probit link}& \multicolumn{4}{c}{Multivariate logit link}\\
                             \cmidrule(lr){3-6}\cmidrule(lr){7-10}\\& True &Mean Est&APB&\multicolumn{1}{p{1.4cm}}{\centering Mean Asym SE}&SD Sample&Mean Est&APB&\multicolumn{1}{p{1.4cm}}{\centering Mean Asym SE}&SD Sample\\ \midrule
$\theta_{1,1}$ & $-$3.0000 & $-$2.9928 & 0.24\% & 0.0369 & 0.0721 & $-$2.9696 & 1.01\% & 0.0469 & 0.0915 \\ 
  $\theta_{1,2}$ & 0.0000 & 0.0018 & \multicolumn{1}{c}{\quad -} & 0.0201 & 0.0655 & 0.0028 & \multicolumn{1}{c}{\quad -} & 0.0210 & 0.0809 \\ 
  $\theta_{1,3}$ & 3.0000 & 2.9966 & 0.11\% & 0.0365 & 0.0685 & 2.9804 & 0.65\% & 0.0467 & 0.0867 \\ 
  $\theta_{2,1}$ & $-$2.0000 & $-$1.9939 & 0.31\% & 0.0292 & 0.0652 & $-$1.9818 & 0.91\% & 0.0355 & 0.0825 \\ 
  $\theta_{2,2}$ & 0.0000 & 0.0013 & \multicolumn{1}{c}{\quad -} & 0.0205 & 0.0634 & 0.0017 & \multicolumn{1}{c}{\quad -} & 0.0205 & 0.0802 \\ 
  $\theta_{2,3}$ & 2.0000 & 1.9979 & 0.10\% & 0.0294 & 0.0706 & 1.9922 & 0.39\% & 0.0360 & 0.0858 \\ 
  $\theta_{3,1}$ & 3.0000 & 3.0006 & 0.02\% & 0.0399 & 0.0738 & 2.9835 & 0.55\% & 0.0490 & 0.0931 \\ 
  $\beta_{1}$ & 2.0000 & 1.9959 & 0.21\% & 0.0225 & 0.0174 & 1.9797 & 1.01\% & 0.0302 & 0.0317 \\ 
  $\beta_{2}$ & $-$1.0000 & $-$0.9993 & 0.07\% & 0.0152 & 0.0120 & $-$0.9892 & 1.08\% & 0.0191 & 0.0222 \\ 
  $\rho_{1,2}$ & 0.9500 & 0.9458 & 0.44\% & 0.0054 & 0.0039 & 0.9495 & 0.05\% & 0.0053 & 0.0036 \\ 
  $\rho_{1,3}$ & 0.8750 & 0.8580 & 1.95\% & 0.0158 & 0.0192 & 0.8768 & 0.21\% & 0.0178 & 0.0107 \\ 
  $\rho_{2,3}$ & 0.8000 & 0.7888 & 1.39\% & 0.0207 & 0.0204 & 0.8007 & 0.09\% & 0.0307 & 0.0139 \\ 
  $\psi_{1}$ & 0.2000 & 0.1917 & 4.15\% & 0.0221 & 0.0219 & 0.2000 & 0.02\% & 0.0284 & 0.0171 \\ 
  $\psi_{2}$ & 0.2500 & 0.2448 & 2.06\% & 0.0219 & 0.0165 & 0.2449 & 2.03\% & 0.0276 & 0.0162 \\ 
  $\psi_{3}$ & 0.3500 & 0.3529 & 0.82\% & 0.0484 & 0.0467 & 0.3583 & 2.37\% & 0.0632 & 0.0374 \\ 
   \bottomrule
\end{tabular}
\end{table}
\begin{table}[ht]
\centering
\caption{This table presents simulation results based on 100 repetitions, $n=1000$, $T = 10$, $q=3$ for the correlation setting $\Sigma$ low and $\Psi$ high.} 
\label{tab:sim_t_10_logit_probit_id_3}
\begin{tabular}{rrrlrrrlrr}
  \toprule
  & & \multicolumn{4}{c}{Multivariate probit link}& \multicolumn{4}{c}{Multivariate logit link}\\
                             \cmidrule(lr){3-6}\cmidrule(lr){7-10}\\& True &Mean Est&APB&\multicolumn{1}{p{1.4cm}}{\centering Mean Asym SE}&SD Sample&Mean Est&APB&\multicolumn{1}{p{1.4cm}}{\centering Mean Asym SE}&SD Sample\\ \midrule
$\theta_{1,1}$ & $-$3.0000 & $-$2.9846 & 0.51\% & 0.0582 & 0.0847 & $-$2.9710 & 0.97\% & 0.0673 & 0.1217 \\ 
  $\theta_{1,2}$ & 0.0000 & 0.0053 & \multicolumn{1}{c}{\quad -} & 0.0438 & 0.0729 & 0.0133 & \multicolumn{1}{c}{\quad -} & 0.0413 & 0.1086 \\ 
  $\theta_{1,3}$ & 3.0000 & 3.0014 & 0.05\% & 0.0583 & 0.0789 & 2.9959 & 0.14\% & 0.0671 & 0.1163 \\ 
  $\theta_{2,1}$ & $-$2.0000 & $-$1.9852 & 0.74\% & 0.0570 & 0.0745 & $-$1.9901 & 0.50\% & 0.0604 & 0.1112 \\ 
  $\theta_{2,2}$ & 0.0000 & 0.0090 & \multicolumn{1}{c}{\quad -} & 0.0521 & 0.0751 & 0.0016 & \multicolumn{1}{c}{\quad -} & 0.0487 & 0.1134 \\ 
  $\theta_{2,3}$ & 2.0000 & 2.0022 & 0.11\% & 0.0571 & 0.0749 & 1.9861 & 0.69\% & 0.0603 & 0.1126 \\ 
  $\theta_{3,1}$ & 3.0000 & 2.9981 & 0.06\% & 0.0865 & 0.1015 & 2.9820 & 0.60\% & 0.0875 & 0.1527 \\ 
  $\beta_{1}$ & 2.0000 & 1.9953 & 0.23\% & 0.0260 & 0.0228 & 1.9867 & 0.67\% & 0.0325 & 0.0317 \\ 
  $\beta_{2}$ & $-$1.0000 & $-$0.9976 & 0.24\% & 0.0175 & 0.0150 & $-$0.9956 & 0.44\% & 0.0204 & 0.0252 \\ 
  $\rho_{1,2}$ & 0.1000 & 0.0938 & 6.21\% & 0.0297 & 0.0231 & 0.1003 & 0.27\% & 0.0354 & 0.0301 \\ 
  $\rho_{1,3}$ & 0.2000 & 0.1993 & 0.37\% & 0.0369 & 0.0308 & 0.1959 & 2.07\% & 0.0501 & 0.0334 \\ 
  $\rho_{2,3}$ & 0.3000 & 0.2991 & 0.29\% & 0.0348 & 0.0332 & 0.2973 & 0.89\% & 0.0474 & 0.0319 \\ 
  $\psi_{1}$ & 0.8000 & 0.7986 & 0.17\% & 0.0079 & 0.0066 & 0.7951 & 0.62\% & 0.0098 & 0.0075 \\ 
  $\psi_{2}$ & 0.8500 & 0.8482 & 0.21\% & 0.0062 & 0.0055 & 0.8442 & 0.68\% & 0.0075 & 0.0068 \\ 
  $\psi_{3}$ & 0.9000 & 0.8986 & 0.15\% & 0.0065 & 0.0061 & 0.8973 & 0.31\% & 0.0085 & 0.0064 \\ 
   \bottomrule
\end{tabular}
\end{table}
\begin{table}[ht]
\centering
\caption{This table presents simulation results based on 100 repetitions, $n=1000$, $T = 10$, $q=3$ for the correlation setting $\Sigma$ low and $\Psi$ low.} 
\label{tab:sim_t_10_logit_probit_id_4}
\begin{tabular}{rrrlrrrlrr}
  \toprule
  & & \multicolumn{4}{c}{Multivariate probit link}& \multicolumn{4}{c}{Multivariate logit link}\\
                             \cmidrule(lr){3-6}\cmidrule(lr){7-10}\\& True &Mean Est&APB&\multicolumn{1}{p{1.4cm}}{\centering Mean Asym SE}&SD Sample&Mean Est&APB&\multicolumn{1}{p{1.4cm}}{\centering Mean Asym SE}&SD Sample\\ \midrule
$\theta_{1,1}$ & $-$3.0000 & $-$2.9849 & 0.50\% & 0.0353 & 0.0648 & $-$2.9704 & 0.99\% & 0.0446 & 0.0933 \\ 
  $\theta_{1,2}$ & 0.0000 & 0.0029 & \multicolumn{1}{c}{\quad -} & 0.0202 & 0.0678 & 0.0026 & \multicolumn{1}{c}{\quad -} & 0.0211 & 0.0803 \\ 
  $\theta_{1,3}$ & 3.0000 & 2.9992 & 0.03\% & 0.0351 & 0.0702 & 2.9822 & 0.59\% & 0.0444 & 0.0877 \\ 
  $\theta_{2,1}$ & $-$2.0000 & $-$1.9884 & 0.58\% & 0.0276 & 0.0645 & $-$1.9776 & 1.12\% & 0.0324 & 0.0807 \\ 
  $\theta_{2,2}$ & 0.0000 & 0.0042 & \multicolumn{1}{c}{\quad -} & 0.0206 & 0.0632 & 0.0027 & \multicolumn{1}{c}{\quad -} & 0.0205 & 0.0763 \\ 
  $\theta_{2,3}$ & 2.0000 & 1.9954 & 0.23\% & 0.0276 & 0.0662 & 1.9852 & 0.74\% & 0.0326 & 0.0841 \\ 
  $\theta_{3,1}$ & 3.0000 & 2.9886 & 0.38\% & 0.0403 & 0.0758 & 2.9787 & 0.71\% & 0.0486 & 0.0884 \\ 
  $\beta_{1}$ & 2.0000 & 1.9933 & 0.34\% & 0.0187 & 0.0169 & 1.9808 & 0.96\% & 0.0254 & 0.0258 \\ 
  $\beta_{2}$ & $-$1.0000 & $-$0.9959 & 0.41\% & 0.0125 & 0.0109 & $-$0.9896 & 1.04\% & 0.0156 & 0.0171 \\ 
  $\rho_{1,2}$ & 0.1000 & 0.1020 & 2.03\% & 0.0178 & 0.0155 & 0.0994 & 0.63\% & 0.0229 & 0.0141 \\ 
  $\rho_{1,3}$ & 0.2000 & 0.1991 & 0.47\% & 0.0346 & 0.0295 & 0.1954 & 2.30\% & 0.0427 & 0.0274 \\ 
  $\rho_{2,3}$ & 0.3000 & 0.2987 & 0.45\% & 0.0345 & 0.0300 & 0.2958 & 1.40\% & 0.0445 & 0.0241 \\ 
  $\psi_{1}$ & 0.2000 & 0.1979 & 1.06\% & 0.0328 & 0.0249 & 0.2039 & 1.94\% & 0.0407 & 0.0261 \\ 
  $\psi_{2}$ & 0.2500 & 0.2398 & 4.09\% & 0.0293 & 0.0245 & 0.2425 & 3.01\% & 0.0361 & 0.0295 \\ 
  $\psi_{3}$ & 0.3500 & 0.1889 & 46.03\% & 0.0709 & 0.2614 & 0.3555 & 1.56\% & 0.0859 & 0.0480 \\ 
   \bottomrule
\end{tabular}
\end{table}
For all scenarios we recover well the true parameters of the proposed model when 
using the probit or logit link, with absolute percentage biases below 1\%
for the threshold and coefficient parameters, with negligible differences to be 
observed between the different correlation scenarios. Higher APBs are observed 
for the correlation parameters, especially for the cases where the true
correlations are low. In Figure~\ref{fig:sim_big_all} one can observe that, 
for the probit link, there is a sign-switch for the estimates of the error
structure in several repetitions or the scenarios $\Sigma$ high -- $\Psi$ low and
$\Sigma$ low -- $\Psi$ low. A closer inspection reveals that this happens
for the persistence parameter $\psi_3$ of the binary response and for $\rho_{13}$
and $\rho_{23}$, which are the cross-correlations for the ordinal outcomes with
the binary outcomes. This issue seems to arise due to the fact that the distribution 
of the binary response is highly imbalanced for those scenarios (see Figure~\ref{fig:sim_big_all}).
For the data application, this can imply that a less imbalanced distribution of the binary response
is desirable, especially if the estimated sign is not intuitive in the 
application context. This can be achieved by e.g.,~by up-sampling.

Mean asymptotic standard errors and sample standard errors are similar in
magnitude for the parameters of the error structure and for the
regression parameters, but we can observe that  
for the threshold parameters
the asymptotic standard errors are overoptimistic. 
This need not be a crucial issue, as typically inference on the threshold 
parameters is not of interest in most applications.
It is to be noted that the 
number of repetitions in this exercise is not particularly large. 
However, this finding is in line with previous observations in the literature 
on composite likelihood methods for longitudinal data \citep{Varin09} and is likely 
due to the instability in estimating the variability matrix $V$.

Computations for this exercise have been performed on 25 IBM dx360M3 nodes 
within a cluster of workstations.

\subsection{Different values of lag parameter $c$}\label{sec:sim_cc}
As mentioned in Section~\ref{sec:inference}, the lag value $c$ determines the 
maximal lag of the longitudinal observations to be included in the pairwise likelihood. 
The main rationale behind using $c < T-1$ is for reducing the computational load
when estimating the parameters. Another aspect is the efficiency of the estimates,
which can suffer when too many pairs are added to the pairwise likelihood 
(see discussion in Section~\ref{sec:inference}).

Based on recommendations in the literature, $c$ can be chosen by minimizing the 
generalized variance (i.e.,~determinant of the covariance matrix of the estimates).  
We wish to investigate in a second simulation exercise whether this approach
performs well for the proposed model.  For this purpose we generate 100 data 
sets from the model with $n=200$ subjects, $T=10$ time points and $q=3$
ordinal responses. As in the previous exercise, we simulate two covariates 
from a standard normal distribution ($p = 2$), which vary for each of the $T$ 
time points, but do not vary with the ordinal outcomes. The vector of regression 
coefficients and the threshold parameters are chosen as in Section~\ref{sec:sim_big}.
For the matrices $\Sigma$ and $\Psi$ we consider a mix of high, medium and low correlation values: 
\begin{align*}
  \Sigma =
  \begin{pmatrix}
    1.000 & 0.100 & 0.500\\
    0.100 & 1.000 & 0.900\\
    0.500 & 0.900 & 1.000
  \end{pmatrix},&\qquad
  \Psi_\text{low} =
    \begin{pmatrix}
    0.800 & 0& 0\\
    0 & 0.500 & 0\\
    0 & 0 & 0.200
  \end{pmatrix}.
\end{align*}
We only perform the exercise for the probit link.
\begin{figure}
\centering
\includegraphics{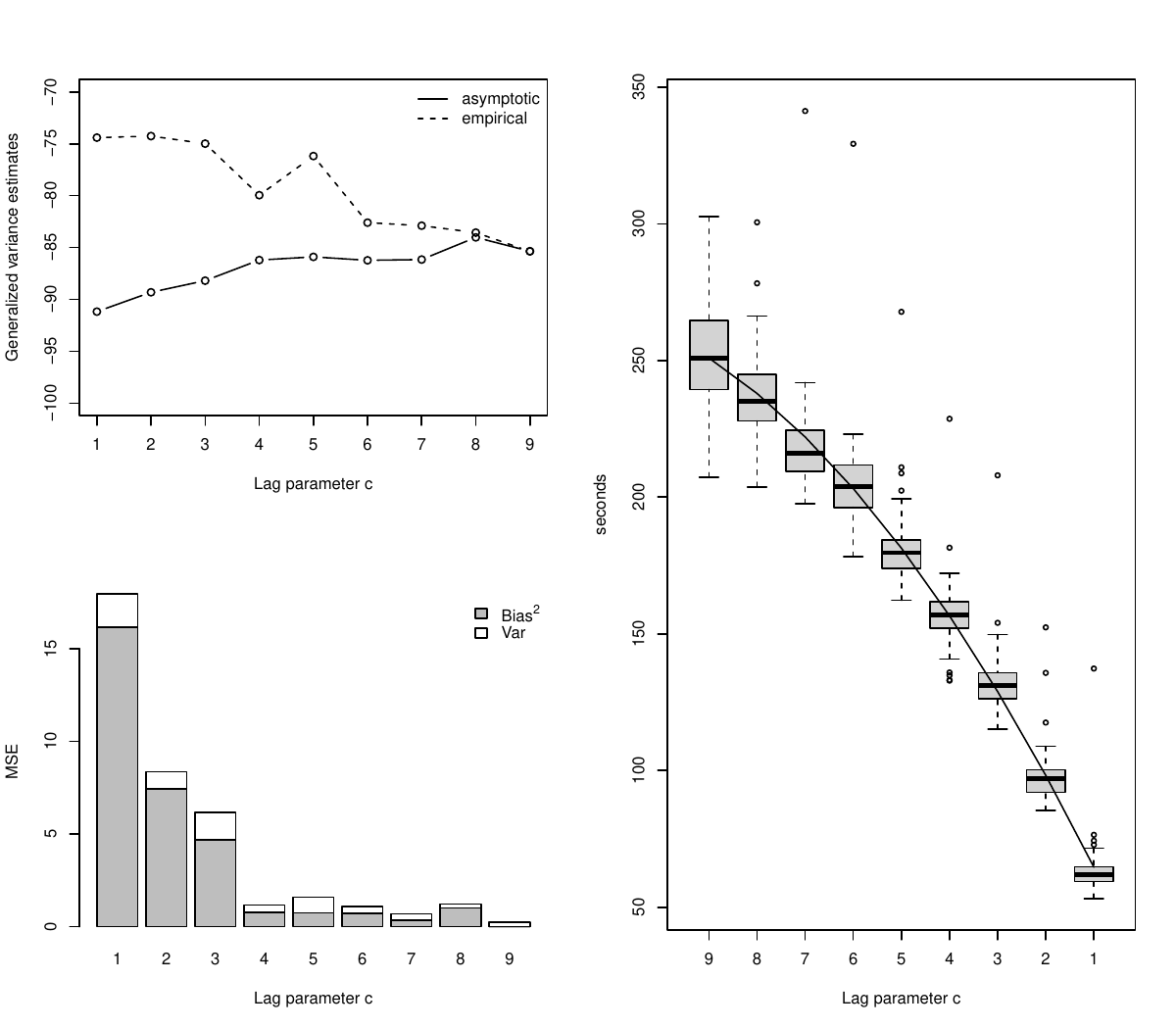}
\caption{The upper left panel  displays the average generalized asymptotic 
variance $\hat g$ (solid line) as well as the empirical covariance of the estimates
in the proposed model (dashed line) over 100 simulated data sets. 
The lower left panel shows the MSE for different lag parameter $c$.
The right panel shows the distribution of the computation time in seconds in the
100 replications for different values of $c$. 
The simulation setting employed contained
$n=200$ subjects, $T = 10$ time points and $q=3$ responses 
(two ordinal responses with four classes and a binary response).}
\label{fig:sim_cc_genvar_rmse_cputime}
\end{figure}
Figure~\ref{fig:sim_cc_genvar_rmse_cputime} summarizes the results of this exercise.
In the upper left panel we show the average estimated log generalized asymptotic variance of the estimates
over $M=100$ simulated data sets. 
\begin{align*}
\hat g(c)= \frac{1}{M}\sum_{m=1}^M\log \det(\hat
H(\hat{\bm\delta}^{(m)}_{PL}(c),c)^{-1}\hat V(\hat{\bm\delta}^{(m)}_{PL}(c),c)
\hat H(\hat{\bm\delta}^{(m)}_{PL}(c),c)^{-1})
\end{align*}
where $\hat{\bm\delta}^{(m)}_{PL}(c)$ denotes the maximum pairwise 
likelihood estimates in replication~$m$ for a fixed value of $c$. We observe that for the 
investigated setting, $\hat g(c)$ is minimal at lag one. 
We superimpose the log determinant of the empirical covariance of the 
estimates in the $M=100$ repetitions in dashed line. We observe that the empirical
and asymptotic generalized variance start converging to each other around lag 8.

In the lower left panel of Figure~\ref{fig:sim_cc_genvar_rmse_cputime}
we show the mean squared error (MSE) for the parameters over 
the $M=100$ replications
\begin{align*}
MSE(c)=\underbrace{(\bar{\bm\delta}_{PL}(c)-\bm\delta)^\top(\bar{\bm\delta}_{PL}(c)-\bm\delta)}_{\text{bias}^2} +
\underbrace{\frac{1}{M}\sum_{m=1}^M(\hat{\bm\delta}^{(m)}_{PL}(c)-\bar{\bm\delta}_{PL}(c))^\top(\hat{\bm\delta}^{(m)}_{PL}(c)-\bar{\bm\delta}_{PL}(c))}_{\text{variance}},\qquad \bar{\bm\delta}_{PL}(c)=\frac{1}{M}\sum_{m=1}^M\hat{\bm\delta}^{(m)}_{PL}(c)
\end{align*}

together with the decomposition in squared bias and variance.
We observe that the MSE decreases only after lag 3 and is minimal at lag 9.
Hence, we find that the generalized variance is not a satisfactory 
measure to select the lag parameter $c$, especially for lower lag values and
that this measure should be employed with care in an empirical application.
A bootstrapping exercise would be therefore beneficial in a real data example,
if lag selection is desired. This can however amount to a significant 
computational burden. Such an exercise would be advisable when 
many model specifications should be compared to each other or when performing
variable selection based on information criteria. The lag would be chosen on the 
most complex model and would be then used for all simpler models 
(the reason being that for model comparison, the pairwise likelihood
of the different models should include the same pairs of observations). 

We also provide the computation time need for the different lags in 
Figure~\ref{fig:sim_cc_genvar_rmse_cputime}. We observe a quadratic relation between the computation time
and the lag parameter $c$. This is explained by the fact that decreasing 
$c$ by 1 results in a reduction in the number of pairs in the likelihood of
$q ^ 2 (T-c-1)(T-c)/2$. 

In spite of the reduction in computation time, based on the MSEs and the empirical
covariance it is suggested that for the studied case reducing $c$ is not connected 
to benefits in terms of inference.

For this exercise we used the solver NEWUOA in 
package \pkg{optimx} \citep{jssoptimx} for optimizing the pairwise likelihood. The
simulations were performed on a MacBook Pro with M1 chip and 16 GB RAM.

\subsection{Comparison with a Bayesian approach}
In final simulation exercise we aim to compare the proposed model estimated by 
package \pkg{mvordflex} with an alternative approach. To date, there are no ready to use 
implementations of such a model, but similar models have been proposed which make
use of Bayesian methods for estimation and inference.
We therefore estimate the proposed model with probit link using the Bayesian approach by 
implementing a program in \pkg{rstan} \citep{pkg:rstan}, the \proglang{R} interface to Stan. 
The estimation is performed using Hamiltonian Monte Carlo methods 
\citep[for more details see e.g.,][]{betancourt2015hamiltonian}.

We implement the proposed model in \pkg{rstan}
and perform a small simulation study to compare the time of computation and 
quality of the estimates using \pkg{mvordflex} and  \pkg{rstan}. 

We again generate $M=100$ data sets from the model with $n=200$ subjects, 
$q=3$ responses and $T=5$ time points.
We keep the same parameter values as Section~\ref{sec:sim_cc}.

\begin{table}[ht]
\centering
\caption{This table presents simulation results based on 100 repetitions, $n=200$, $T=5$, $q=3$ for the proposed model where the estimation using Bayesian methods is compared to the pairwise likelihood estimation.} 
\label{tab:sim_rstan_mvordflex}
\begin{tabular}{rrrrrrrr}
  \toprule
  & & \multicolumn{3}{c}{Pairwise likelihood (probit link)}& \multicolumn{3}{c}{Bayesian model with rstan}\\
                             \cmidrule(lr){3-5}\cmidrule(lr){6-8}\\& True &Mean Est&APB&SD Sample&Mean Est&APB&SD Sample\\ \midrule
$\theta_{1,1}$ & $-$3.0000 & $-$2.9847 & 0.0051 & 0.1995 & $-$2.9173 & 0.0276 & 0.1578 \\ 
  $\theta_{1,2}$ & 0.0000 & $-$0.0024 &  & 0.1661 & 0.0249 &  & 0.1040 \\ 
  $\theta_{1,3}$ & 3.0000 & 2.9994 & 0.0002 & 0.1959 & 2.9297 & 0.0234 & 0.1703 \\ 
  $\theta_{2,1}$ & $-$2.0000 & $-$1.9757 & 0.0122 & 0.1551 & $-$2.0443 & 0.0222 & 0.1157 \\ 
  $\theta_{2,2}$ & 0.0000 & 0.0009 &  & 0.1434 & 0.0005 &  & 0.0689 \\ 
  $\theta_{2,3}$ & 2.0000 & 1.9899 & 0.0051 & 0.1533 & 2.0387 & 0.0194 & 0.1200 \\ 
  $\theta_{3,1}$ & 3.0000 & 3.0121 & 0.0040 & 0.1565 & 3.0734 & 0.0245 & 0.1467 \\ 
  $\beta_{1}$ & 2.0000 & 1.9900 & 0.0050 & 0.0538 & 2.0013 & 0.0006 & 0.0935 \\ 
  $\beta_{2}$ & $-$1.0000 & $-$1.0007 & 0.0007 & 0.0393 & $-$1.0061 & 0.0061 & 0.0571 \\ 
  $\rho_{1,2}$ & 0.1000 & 0.0997 & 0.0027 & 0.0599 & 0.0888 & 0.1120 & 0.0547 \\ 
  $\rho_{1,3}$ & 0.5000 & 0.4668 & 0.0664 & 0.1085 & 0.3449 & 0.3101 & 0.1139 \\ 
  $\rho_{2,3}$ & 0.9000 & 0.8918 & 0.0091 & 0.0474 & 0.7935 & 0.1183 & 0.0560 \\ 
  $\psi_{1}$ & 0.8000 & 0.7908 & 0.0115 & 0.0195 & 0.9226 & 0.1533 & 0.0375 \\ 
  $\psi_{2}$ & 0.5000 & 0.4894 & 0.0212 & 0.0489 & 0.6532 & 0.3063 & 0.0602 \\ 
  $\psi_{3}$ & 0.2000 & 0.2050 & 0.0249 & 0.1133 & 0.4143 & 1.0715 & 0.1268 \\ 
   \bottomrule
\end{tabular}
\end{table}
\begin{figure}
\centering
\includegraphics{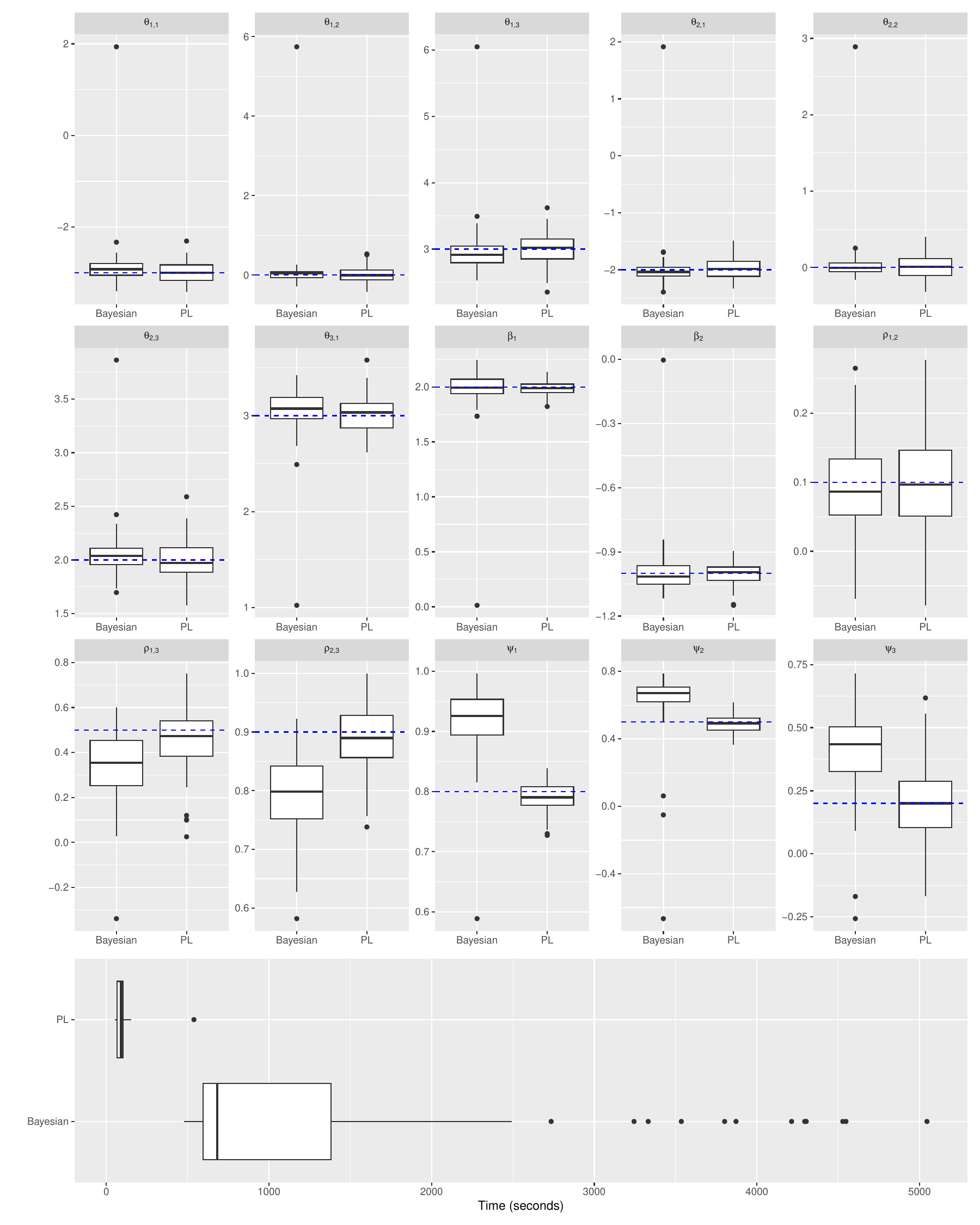}
\caption{This figure displays the distribution for 100 simulated data sets
of the coefficients for the model estimated with pairwise likelihood methods (PL) 
and the posterior means in the Bayesian model estimated with \pkg{rstan}. 
The lower panel shows the distribution of the computation time in seconds.
The simulation setting employed contained
$n=200$ subjects, $T = 5$ time points and $q=3$ responses (two ordinal responses 
with four classes and a binary response).}
\label{fig:sim_mvordflex_rstan}
\end{figure}

In the estimation of the Bayesian model we use one chain with 2000 draws, out of 
which 1000 are discarded after warm-up. 
The average computation across the 100 repetitions is 
90.04 seconds for the model estimated with 
\pkg{mvordflex} and 1335.24 seconds for the Bayesian model.
The bottom panel of Figure~\ref{fig:sim_mvordflex_rstan} 
illustrates the distribution of the computation time in seconds 
over the 100 repetitions.

Figure~\ref{fig:sim_mvordflex_rstan} shows the distribution over the 100 repetitions
of estimated coefficients by \pkg{mvordflex} and the distribution of the 
posterior means of the parameters in the Bayesian model. 
Table~\ref{tab:sim_rstan_mvordflex} provides some
summary statistics. We observe that the
true threshold and coefficient parameters are recovered well by both methods,
while the Bayesian method is performing more poorly for recovering some of the
error structure parameters. It should be noted that generic priors have been used 
in the implementation of the Bayesian model, so further tuning of the priors 
and of the starting values could make the Bayesian model more competitive.
Moreover, the traceplots of the Bayesian model reveal some degree of 
autocorrelation, so increasing the number of draws 
and thinning the Markov chains would lead to more efficient estimates. 
All in all, we show that for the investigated setting, the pairwise likelihood approach
for the proposed model performs well both in terms of parameter estimation and
in terms of computation time.

For this exercise we used the solver \code{nlminb} in 
package \pkg{optimx} \citep{jssoptimx} for optimizing the pairwise likelihood. The
simulations were performed on a MacBook Pro with M1 chip and 16 GB RAM.

\section{Empirical analysis}\label{sec:empirical_results}
We apply the proposed model to corporate credit ratings from
Standard and Poor's (S\&P) and Moody's as well as to a failure information
indicator over the period of 2003--2013. The flexible framework allows to
account for the time persistence of the ratings and failures, as well as for
the correlation among the raters and failure dimensions.

\subsection{Data}\label{sec:data}
In this paper we use S\&P long-term issuer credit ratings from the
Compustat-Capital IQ Credit Ratings database as well as issuer credit ratings
from Moody's. S\&P provides its ratings on a scale with 21
non-default categories ranging from AAA to C. Moody's uses a different scale by
assigning 21 rating classes ranging from Aaa to the default class C. The failure
indicator is constructed based on the default data from the
UCLA-LoPucki Bankruptcy Research Database and the Mergent issuer default file.
A binary failure indicator is constructed in the following way: a default is
recorded in a year if a firm filed for bankruptcy under Chapter 7 or
Chapter 11 or the firm receives a default rating from one of the CRAs in the year
following the rating observation. This definition is similar to definition of
\cite{campbell2008search}, and to the promoted
default definition in \cite{baselii}.

For the construction of the firm-level variables we make use the Compustat
and CRSP databases together with the corresponding linking files available on
Wharton research data services (WRDS). We use the pre-calculated financial
ratios available in the \textit{Financial Ratios Suite}.
We include in the analysis the universe of Compustat/CRSP US
corporates which have at least one S\&P rating observation or a
rating observation of Moody's in the period from
2003 to 2013 $(T=11)$. Following common practice, we exclude financial, utility and real estate firms from
the data set, as their credit risk is different than that of other sectors (mainly
due to the higher likelihood of intervention of the government in case of default).
The end of year ratings are merged to the financial ratios on a
calendar year basis. We perform this by assigning the
latest financial statement available before year endsto the end-of-year ratings.
For the computation of the market variables we use daily stock
price data available an CRSP. Winsorization of all explanatory variables at the 99th percentile as well as
for ratios with negative values at the 1st percentile is conducted.

As explanatory variables we make use of the $p=7$ variables proposed by
\cite{Tian2015}, who select these variables based on a statistical variable
selection exercise and show that improved performance when predicting defaults
in a static setting compared to other established models in the credit risk
literature.
\begin{table}[h!]
\centering
\begin{tabular}{lD{.}{.}{5}D{.}{.}{5}D{.}{.}{5}D{.}{.}{5}D{.}{.}{5}D{.}{.}{5}D{.}{.}{5}}
\toprule
 \multicolumn{1}{c}{\rotatebox{90}{}} & \multicolumn{1}{c}{\rotatebox{90}{SIGMA}} & \multicolumn{1}{c}{\rotatebox{90}{lct/at}} & \multicolumn{1}{c}{\rotatebox{90}{debt/at}} & \multicolumn{1}{c}{\rotatebox{90}{ni/mta}} & \multicolumn{1}{c}{\rotatebox{90}{lt/mta}} & \multicolumn{1}{c}{\rotatebox{90}{PRICECAP}} & \multicolumn{1}{c}{\rotatebox{90}{EXRET}} \\ \midrule
\multicolumn{7}{c}{Entire data set}\\
\midrule
 Min. & 0.0072 & 0.0000 & 0.0000 & -1.1180 & 0.0110 & 1.6601 & -1.9155 \\ 
  1st Qu. & 0.0141 & 0.1220 & 0.1640 & 0.0060 & 0.2530 & 12.5700 & -0.1941 \\ 
  Median & 0.0195 & 0.1910 & 0.2770 & 0.0290 & 0.3970 & 15.0000 & 0.0073 \\ 
  Mean & 0.0228 & 0.2112 & 0.3069 & 0.0109 & 0.4176 & 13.0108 & -0.0139 \\ 
  3rd Qu. & 0.0271 & 0.2770 & 0.4160 & 0.0430 & 0.5630 & 15.0000 & 0.2041 \\ 
  Max. & 0.0908 & 0.8110 & 0.8940 & 0.3000 & 0.9780 & 15.0000 & 1.1837 \\ \midrule
\multicolumn{7}{c}{Failure Group}\\
\midrule
 Min. & 0.0188 & 0.0000 & 0.0340 & -0.5440 & 0.2860 & 1.6601 & -1.9155 \\ 
  1st Qu. & 0.0414 & 0.1320 & 0.3910 & -0.1700 & 0.7920 & 2.4239 & -1.8766 \\ 
  Median & 0.0553 & 0.2160 & 0.5490 & -0.0810 & 0.8940 & 5.2110 & -1.2051 \\ 
  Mean & 0.0597 & 0.2596 & 0.5446 & -0.1201 & 0.8357 & 6.4777 & -1.0975 \\ 
  3rd Qu. & 0.0827 & 0.3530 & 0.6740 & -0.0150 & 0.9330 & 9.2451 & -0.5710 \\ 
  Max. & 0.0908 & 0.7980 & 0.8940 & 0.0390 & 0.9780 & 15.0000 & 1.1837 \\ \bottomrule
\end{tabular}
\caption[Summary statistics]{Summary statistics of all variables for the entire data set and the failure group.}
\label{tab:ratios}
\end{table}
\begin{figure}
\centering
\includegraphics{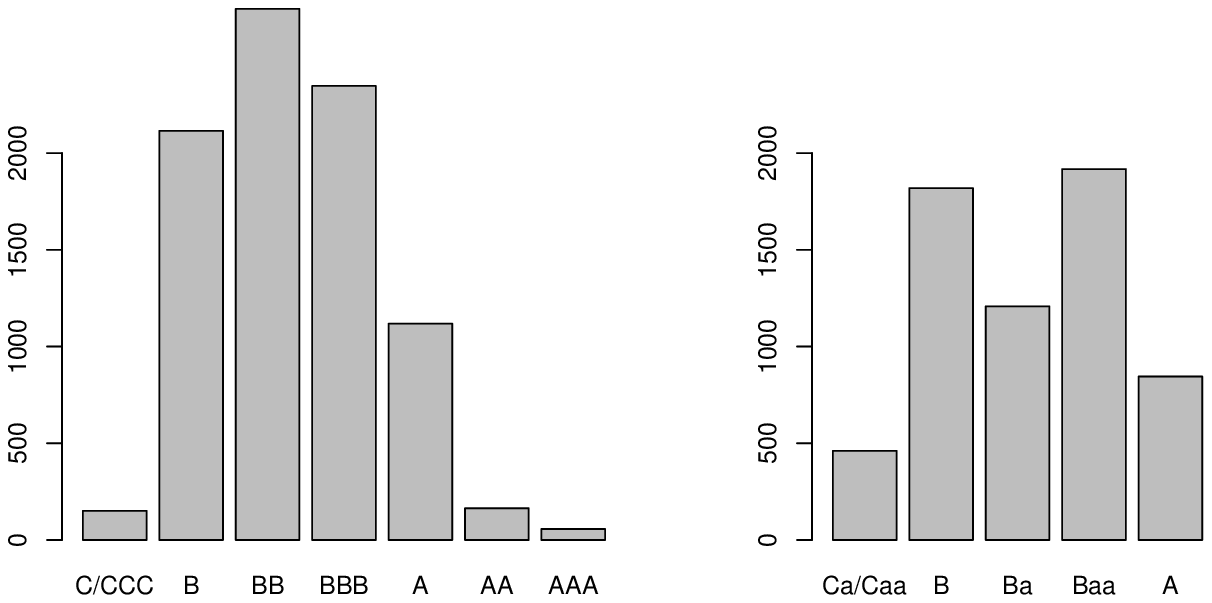}
\caption{Rating distribution over the whole sample period 2003--2013.}
\label{fig:rat_dist}
\end{figure}
Table~\ref{tab:ratios} summarizes the explanatory variables
idiosyncratic risk ($SIGMA$),
short-term liabilities-to-assets ratio ($lct/at$),
debt-to-assets ratio ($debt/at$),
net income-to-market assets ($ni/mta$),
total liabilities-to-market assets ratio ($lt/mta$),
stock price capped at 15\$ \sloppy ($PRICECAP$) and
excess return over the ARCA/AMEX index ($EXRET$) for the entire data set and the
failure group. We observe noticeable higher means and medians for the idiosyncratic risk
$SIGMA$, and for the liability-related ratios $lct/at$, $debt/at$ and $lt/mta$ in the failure group compared to the entire sample,
while firms in the failure group have on average lower $ni/mta$,  $PRICECAP$,
and $EXRET$ values.

In total, the obtained data set comprises
1519 firms with
11277 firm-year observations.
\begin{table}[ht]
\centering
\begin{tabular}{lrrrrrrrrrrrr}
  \toprule
  & 2003 & 2004 & 2005 & 2006 & 2007 & 2008 & 2009 & 2010 & 2011 & 2012 & 2013 & Total \\ 
  \midrule & \multicolumn{12}{c}{Failure distribution in entire sample}\\ \midrule
Fail & 12 & 9 & 5 & 4 & 13 & 32 & 6 & 7 & 3 & 4 & 6 & 101 \\ 
  No Fail & 1124 & 1118 & 1098 & 1090 & 1039 & 986 & 980 & 956 & 932 & 930 & 923 & 11176 \\ 
  Fail. Rate (\%) & 1.06 & 0.80 & 0.45 & 0.37 & 1.24 & 3.14 & 0.61 & 0.73 & 0.32 & 0.43 & 0.65 & 0.90 \\ 
   \midrule & \multicolumn{12}{c}{S\&P rating distribution}\\ \midrule C/CCC & 17 & 14 & 15 & 18 & 12 & 21 & 23 & 6 & 7 & 7 & 11 & 151 \\ 
  B & 186 & 191 & 190 & 205 & 206 & 200 & 198 & 199 & 175 & 188 & 177 & 2115 \\ 
  BB & 294 & 288 & 277 & 282 & 268 & 240 & 211 & 214 & 227 & 219 & 225 & 2745 \\ 
  BBB & 228 & 240 & 226 & 211 & 194 & 194 & 198 & 212 & 207 & 219 & 218 & 2347 \\ 
  A & 119 & 115 & 118 & 108 & 102 & 93 & 90 & 91 & 93 & 92 & 97 & 1118 \\ 
  AA & 18 & 16 & 14 & 14 & 13 & 15 & 15 & 15 & 13 & 14 & 17 & 164 \\ 
  AAA & 7 & 6 & 6 & 6 & 6 & 6 & 4 & 4 & 4 & 4 & 4 & 57 \\ 
   \midrule & \multicolumn{12}{c}{Distribution of failures  among the firms rated by S\&P}\\ \midrule Fail & 11 & 9 & 4 & 4 & 11 & 30 & 6 & 4 & 3 & 1 & 5 & 88 \\ 
  No Fail & 858 & 861 & 842 & 840 & 790 & 739 & 733 & 737 & 723 & 742 & 744 & 8609 \\ 
  Fail. Rate (\%) & 1.27 & 1.03 & 0.47 & 0.47 & 1.37 & 3.90 & 0.81 & 0.54 & 0.41 & 0.13 & 0.67 & 1.01 \\ 
   \midrule & \multicolumn{12}{c}{Moody's rating distribution}\\ \midrule Ca/Caa & 37 & 36 & 31 & 45 & 40 & 55 & 59 & 42 & 38 & 35 & 43 & 461 \\ 
  B & 197 & 184 & 187 & 185 & 173 & 154 & 143 & 151 & 149 & 154 & 142 & 1819 \\ 
  Ba & 113 & 124 & 121 & 113 & 115 & 106 & 101 & 108 & 108 & 98 & 101 & 1208 \\ 
  Baa & 175 & 183 & 171 & 167 & 168 & 166 & 164 & 174 & 171 & 189 & 189 & 1917 \\ 
  A & 90 & 81 & 83 & 82 & 75 & 72 & 69 & 71 & 77 & 73 & 72 & 845 \\ 
  Aa & 13 & 14 & 12 & 13 & 11 & 10 & 10 & 10 & 10 & 10 & 10 & 123 \\ 
  Aaa & 5 & 5 & 5 & 4 & 2 & 2 & 2 & 2 & 2 & 2 & 2 & 33 \\ 
   \midrule & \multicolumn{12}{c}{Distribution of failures  among the firms rated by Moody's}\\ \midrule Fail & 7 & 6 & 3 & 4 & 5 & 17 & 3 & 2 & 2 & 1 & 4 & 54 \\ 
  No Fail & 623 & 621 & 607 & 605 & 579 & 548 & 545 & 556 & 553 & 560 & 555 & 6352 \\ 
  Fail. Rate (\%) & 1.11 & 0.96 & 0.49 & 0.66 & 0.86 & 3.01 & 0.55 & 0.36 & 0.36 & 0.18 & 0.72 & 0.84 \\ 
   \bottomrule
\end{tabular}
\caption{Failure and rating distribution in the sample years.} 
\label{tab:ratdist}
\end{table}Figure~\ref{fig:rat_dist} shows the overall rating distribution for Moody's and S\&P. The rating distribution for  S\&P is unimodal with a mode in the $BB$ rating class. Moody's most
frequent rating class is $Baa$ followed by $B$.
Note that we do not observe all the ratings at all the time points. For
77\% of the observations we observe S\&P ratings, while we have a coverage of
57\% for Moody's.

Finally, Table~\ref{tab:ratdist} contains the failure and rating distribution for the
sample years 2003--2013.
We observe that the sample contains a failure rate of
below 1\%.
\subsection{Model fit}
We fit the proposed multivariate ordinal regression model to the credit risk
data set using the covariates introduced in Section~\ref{sec:data}.
The data set contains three outcomes, namely the S\&P ratings, Moody's ratings
and the failure indicator $(q=3)$.
We make the following parameterization choices: for each response
we fit separate sets of regression coefficients. This is motivated by the fact that
in the literature not all covariates are expected to have the same effect on the
ratings and on the failure dimensions.
Similarly, we fit three different sets of threshold parameters. Due to the high
number of parameters to be estimated, we remove the modifiers in the credit
ratings and obtain a rating scale with each 7 classes for S\&P and Moody's.
Furthermore, the regression coefficients and threshold parameters are assumed
to be constant over time. This gives rise to a contrast matrix $C$ with  $374 = 11 \times (6 + 6 + 1) + 3 \times 11 \times 7$ rows and $34 = 6 + 6 + 1 + 7 \times 3$ columns.
\begin{align*}
C=
\left(\begin{array}{c}
\underbrace{(1,\ldots,1)^\top }_{T=11\, \text{times}}\otimes
  \left(\begin{matrix}
  \diag{I_6, I_6, 1}& \bm 0_{13 \times 21} \\
  \end{matrix}\right)\\
\underbrace{(1,\ldots,1)^\top }_{T=11\, \text{times}}\otimes
  \left(\begin{matrix}
  \bm 0_{21 \times 13}&\diag{I_7, I_7, I_7}\\
  \end{matrix}\right)
\end{array}\right).
\end{align*}

\begin{figure}
\centering
\includegraphics{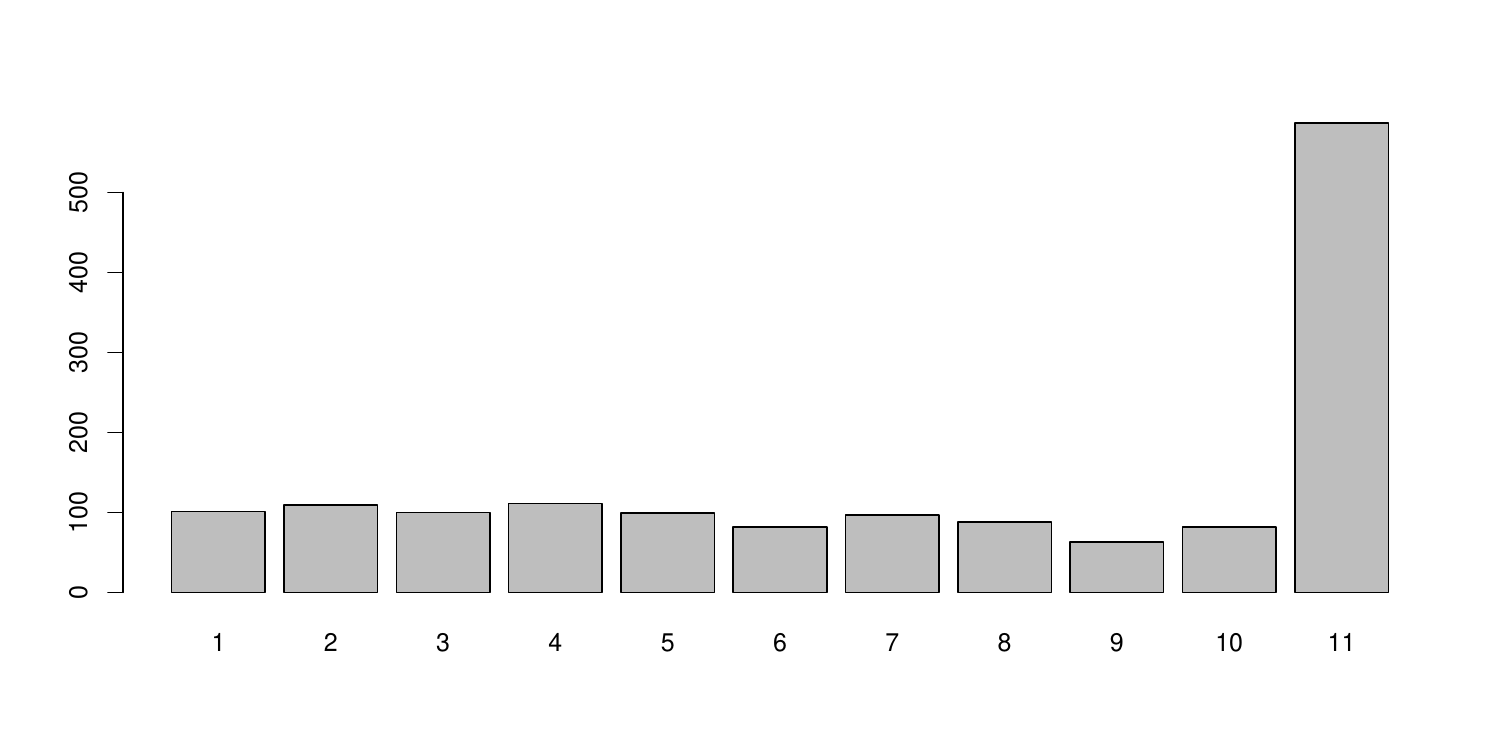}
\caption{Distribution of the number of time points observed per firm in the sample.}
\label{fig:ntimepoints}
\end{figure}

Finally, we motivate the autoregressive lag of order one by the short length of the time series.
Figure~\ref{fig:ntimepoints} distribution of the number of time points observed 
per firm in the sample. Given that the companies can enter (exit) the sample
after (before) the start (end) of the sample period, we observe
587 firms with all time points
observed.
Moreover, for the firms with more than two observed time points we perform the following exercise.
For each of the three responses, we estimate a separate ordinal model with iid errors
using the \pkg{ordinal} \proglang{R} package \citep{pkgordinal}.
For each of the three models we calculate the surrogate residuals proposed in \cite{doi:10.1080/01621459.2017.1292915}
using the \pkg{sure} \proglang{R} package \citep{pkg:sure}.
We then run an ARIMA model on the residuals for each firm and each response
and select the optimal autoregressive lag based on AIC.
This analysis shows that for the large majority of the firms either zero or one
autoregressive lag has been selected, with only a few firms having a selection of
lag two.

Finally, given the low failure rate in the sample, we decided to up-sample the
defaulted companies to have a ratio of 50\% failed vs 50\% non-failed companies
in the sample. Note that this does not lead to a 50\% default rate in the sample,
as the whole history of the defaulted firms is repeatedly included in the sample.
After up-sampling, we achieve an overall failure rate of
11.98\%.

We estimate this model specification with both the logit and probit link and
find the model with logit link to have a better fit based on the composite likelihood Akaike information criterion (979045.65 logit link
 vs. 990175.88 probit link; note that a lower value implies
 a better fit) and composite likelihood Bayesian information criterion (993631.26 logit link
 vs. 1004452.16 probit link).
The estimated parameters of the model with logit link are shown in
the first column of Table~\ref{tab:full_comparison}.
%
%


The model provides information on the difference in the covariates among the
three outcomes: S\&P ratings, Moody's ratings and the failure indicator.
Given that both the threshold and the regression parameters are allowed to vary
with the outcome, direct comparisons among the magnitude of
coefficients and thresholds of the different outcomes have to be
performed with care. This is due to the fact that in ordinal models absolute
location and absolute scale are not identifiable.
Finally, the proposed error structure gives insights into the time-persistence
of the separate outcomes as well as the contemporaneous correlations among the three outcomes.

When analyzing the regression coefficients displayed in
Table~\ref{tab:full_comparison},
we find that the signs for most of the significant coefficients are as expected.
The variables  $lct/at$, $debt/at$ and $ni/mta$ are non-significant for the
failure dimensions at a 5\% significance level.
The excess return $EXRET$ variable has a positive sign, meaning that higher
excess returns lead to higher values in the default process.
On the other hand, this variable has
a negative sign for the rating dimensions, which can imply that
in the rating process
firms with high excess returns can be considered riskier.

Qualitatively, the coefficients of the two rating dimensions are rather similar,
so we can assume that they have a similar scale. This allows us to have a look
at the threshold parameters for these two outcomes
and cautiously make some interpretations.
As the results in the first column of Table~\ref{tab:full_comparison} show, we observe lower thresholds
estimated for S\&P compared to Moody's in the speculative grades. This can
translate into Moody's being more conservative in the speculative grade regions.
Note that the differences in the investment grade categories are negligible.

The estimated error structure provides information on the inter-rater dependence
as well as on the time persistence of the ratings. The estimated parameters of
the error structure can be found in the bottom part of the first column in Table~\ref{tab:full_comparison}. As expected, we find a high
correlation of 0.92
among the S\&P and Moody's ratings and  a lower correlation between the raters
and the failure indicator. We observe a correlation of
0.3
between S\&P and the failure dimension and a correlation of
0.33
between Moody's and the failure dimension.
The second component of the error structure provides information on the time
persistence for each dimension separately.
The time persistence is rather high all outcomes.
We observe for the time persistence parameter $\xi$ an estimated value of
0.93
for S\&P,
0.94
for Moody's and
0.77
for the failure component.


\subsection{Model comparison}
When comparing the panel data model with simpler models based on the composite
likelihood AIC we observe that the proposed model
(AIC: 979045.65,
 BIC~993631.26)
is indeed preferred over simpler specifications.
We consider a model which takes no correlation into account
(AIC: 1038281.48,
 BIC: 1053091.83),
a model which takes only the cross-sectional dependence into account
(AIC: 1031555.63,
 BIC: 1046438.48)
 and a model which takes only the  longitudinal dependence into account
(AIC: 1007943.9,
 BIC: 1022511.19).
It can be observed that the longitudinal specification improves the fit more
than the cross-sectional specification.
\begin{table}[ht]
\centering
\begin{tabular}{lrrrr}
  \toprule
  Parameter & Full & Cross-sectional &Longitudinal & i.i.d. \\
  \midrule
$\beta_{\text{SIGMA, S\&P}}$ & $-$144.0603 (1.3252)*** & $-$68.8869 (1.6678)*** & $-$150.9240 (1.4198)*** & $-$69.2209 (1.6534)*** \\ 
  $\beta_{\text{SIGMA, M}}$ & $-$144.6722 (1.4365)*** & $-$67.8798 (1.9098)*** & $-$152.3875 (1.6251)*** & $-$68.4400 (1.9226)*** \\ 
  $\beta_{\text{SIGMA, Fail}}$ &  $-$21.7672 (1.5487)*** & $-$14.7220 (1.5684)*** &  $-$20.5287 (1.5693)*** & $-$14.5111 (1.5556)*** \\ 
  $\beta_{\text{lct/at, S\&P}}$ &    3.3701 (0.1962)*** &   1.3313 (0.1809)*** &    3.2531 (0.1831)*** &   1.3315 (0.1775)*** \\ 
  $\beta_{\text{lct/at, M}}$ &    3.5661 (0.2500)*** &   1.2803 (0.2410)*** &    3.2786 (0.2598)*** &   1.2850 (0.2425)*** \\ 
  $\beta_{\text{lct/at, Fail}}$ &   $-$0.0858 (0.2889)\phantom{***} &  $-$0.0726 (0.2813)\phantom{***} &   $-$0.1393 (0.2943)\phantom{***} &  $-$0.0883 (0.2819)\phantom{***} \\ 
  $\beta_{\text{debt/at, S\&P}}$ &   $-$5.5743 (0.1470)*** &  $-$2.0501 (0.1453)*** &   $-$5.1402 (0.1475)*** &  $-$2.0512 (0.1443)*** \\ 
  $\beta_{\text{debt/at, M}}$ &   $-$5.8692 (0.1757)*** &  $-$2.0287 (0.1810)*** &   $-$5.3483 (0.1881)*** &  $-$2.0473 (0.1821)*** \\ 
  $\beta_{\text{debt/at, Fail}}$ &   $-$0.3459 (0.1959).\phantom{.**} &  $-$0.1948 (0.1908)\phantom{***} &   $-$0.2827 (0.1983)\phantom{***} &  $-$0.1977 (0.1908)\phantom{***} \\ 
  $\beta_{\text{ni/mta, S\&P}}$ &    3.2805 (0.1291)*** &   1.6967 (0.1972)*** &    3.6080 (0.1470)*** &   1.6899 (0.1951)*** \\ 
  $\beta_{\text{ni/mta, M}}$ &    4.9264 (0.1731)*** &   2.4528 (0.2876)*** &    5.3040 (0.2179)*** &   2.5055 (0.2986)*** \\ 
  $\beta_{\text{ni/mta, Fail}}$ &    0.2418 (0.2193)\phantom{***} &   0.1630 (0.2230)\phantom{***} &    0.2542 (0.2218)\phantom{***} &   0.1698 (0.2223)\phantom{***} \\ 
  $\beta_{\text{lt/mta, S\&P}}$ &   $-$5.1035 (0.1466)*** &  $-$1.9668 (0.1447)*** &   $-$4.9137 (0.1451)*** &  $-$1.9597 (0.1427)*** \\ 
  $\beta_{\text{lt/mta, M}}$ &   $-$6.1562 (0.1777)*** &  $-$2.2461 (0.1890)*** &   $-$5.7101 (0.1893)*** &  $-$2.2559 (0.1895)*** \\ 
  $\beta_{\text{lt/mta, Fail}}$ &   $-$4.4524 (0.2188)*** &  $-$2.9063 (0.2128)*** &   $-$4.1215 (0.2210)*** &  $-$2.9058 (0.2126)*** \\ 
  $\beta_{\text{PRICECAP, S\&P}}$ &    0.2453 (0.0059)*** &   0.0990 (0.0068)*** &    0.2410 (0.0062)*** &   0.1000 (0.0068)*** \\ 
  $\beta_{\text{PRICECAP, M}}$ &    0.2249 (0.0072)*** &   0.0828 (0.0089)*** &    0.2115 (0.0083)*** &   0.0828 (0.0090)*** \\ 
  $\beta_{\text{PRICECAP, Fail}}$ &    0.0880 (0.0092)*** &   0.0523 (0.0090)*** &    0.0782 (0.0093)*** &   0.0521 (0.0090)*** \\ 
  $\beta_{\text{EXRET, S\&P}}$ &   $-$0.3934 (0.0198)*** &  $-$0.1427 (0.0268)*** &   $-$0.3798 (0.0215)*** &  $-$0.1416 (0.0266)*** \\ 
  $\beta_{\text{EXRET, M}}$ &   $-$0.6003 (0.0229)*** &  $-$0.1884 (0.0317)*** &   $-$0.5118 (0.0261)*** &  $-$0.1861 (0.0319)*** \\ 
  $\beta_{\text{EXRET, Fail}}$ &    1.2890 (0.0370)*** &   0.8831 (0.0385)*** &    1.2061 (0.0378)*** &   0.8854 (0.0386)*** \\ 
   \midrule $\theta_{\text{C/CCC|B}}$ & $-$16.2744 (0.1746)*** & $-$6.6801 (0.1833)*** & $-$15.9077 (0.1771)*** & $-$6.6970 (0.1822)*** \\ 
  $\theta_{\text{B|BB}}$ &  $-$6.9537 (0.1425)*** & $-$2.7923 (0.1531)*** &  $-$6.6705 (0.1464)*** & $-$2.7794 (0.1521)*** \\ 
  $\theta_{\text{BB|BBB}}$ &  $-$3.0256 (0.1354)*** & $-$1.3118 (0.1471)*** &  $-$2.9805 (0.1394)*** & $-$1.2962 (0.1459)*** \\ 
  $\theta_{\text{BBB|A}}$ &   0.5283 (0.1357)*** & $-$0.0143 (0.1453)\phantom{***} &   0.2962 (0.1384)*\phantom{**} &  0.0004 (0.1442)\phantom{***} \\ 
  $\theta_{\text{A|AA}}$ &   4.6396 (0.1733)*** &  1.4564 (0.1738)*** &   4.0619 (0.1728)*** &  1.4711 (0.1729)*** \\ 
  $\theta_{\text{AA|AAA}}$ &   7.6309 (0.2600)*** &  2.5069 (0.2510)*** &   6.7974 (0.2580)*** &  2.5225 (0.2506)*** \\ 
  $\theta_{\text{Ca/Caa|B}}$ & $-$12.4197 (0.1874)*** & $-$4.7612 (0.2106)*** & $-$11.7874 (0.2029)*** & $-$4.7972 (0.2120)*** \\ 
  $\theta_{\text{B|Ba}}$ &  $-$6.0949 (0.1707)*** & $-$2.4280 (0.1951)*** &  $-$5.9053 (0.1879)*** & $-$2.4505 (0.1965)*** \\ 
  $\theta_{\text{Ba|Baa}}$ &  $-$3.6720 (0.1679)*** & $-$1.5850 (0.1937)*** &  $-$3.7332 (0.1854)*** & $-$1.6071 (0.1950)*** \\ 
  $\theta_{\text{Baa|A}}$ &   0.3802 (0.1654)*\phantom{**} & $-$0.2066 (0.1888)\phantom{***} &  $-$0.1532 (0.1811)\phantom{***} & $-$0.2269 (0.1901)\phantom{***} \\ 
  $\theta_{\text{A|Aa}}$ &   4.8088 (0.2078)*** &  1.2795 (0.2197)*** &   3.7642 (0.2185)*** &  1.2653 (0.2212)*** \\ 
  $\theta_{\text{Aa|Aaa}}$ &   8.4246 (0.3219)*** &  2.4827 (0.3238)*** &   6.9670 (0.3235)*** &  2.4747 (0.3234)*** \\ 
  $\theta_{\text{Fail|NoFail}}$ &  $-$5.3054 (0.2409)*** & $-$3.4748 (0.2367)*** &  $-$4.8987 (0.2438)*** & $-$3.4732 (0.2367)*** \\ 
   \midrule $\rho$ S\&P M & 0.9199 (0.0054)*** & 0.9308 (0.0037)*** &  &  \\ 
  $\rho$ S\&P Fail & 0.2979 (0.0299)*** & 0.2609 (0.0225)*** &  &  \\ 
  $\rho$ M Fail & 0.3252 (0.0468)*** & 0.2379 (0.0339)*** &  &  \\ 
  $\xi$ S\&P & 0.9317 (0.0028)*** &  & 0.9190 (0.0027)*** &  \\ 
  $\xi$ M & 0.9421 (0.0028)*** &  & 0.9235 (0.0030)*** &  \\ 
  $\xi$ Fail & 0.7692 (0.0103)*** &  & 0.7150 (0.0111)*** &  \\ 
   \midrule CLAIC & 979045.652 & 1031555.629 & 1007943.897 & 1038281.479 \\ 
  CLBIC & 993631.264 & 1046438.478 & 1022511.192 & 1053091.828 \\ 
   \bottomrule\end{tabular}
\caption{Estimated parameters for four different models: the full three-dimensional model, the cross-sectional model, the longitudinal model which assumes an AR(1) error structure on each ordinal outcome and the model where the observations are considered i.i.d both in time and in cross-section. Significance codes: *** -- $p$-value < 0.001, ** -- $p$-value $\in [0.001, 0.01)$, * -- $p$-value $\in [0.01, 0.05)$,  . --  $p$-value $\in [0.05, 0.1)$.}
\label{tab:full_comparison}
\end{table}
Table~\ref{tab:full_comparison} presents the parameter estimates for the different models.
Note again that the absolute value of the coefficients is not directly comparable, given
the non-identifiability of the scale.


\section{Software implementation}\label{sec:software}
In this section we illustrate how the model proposed above can be estimated using
the \proglang{R} implementation in package \pkg{mvordflex}. This package is designed as an extension to the \pkg{mvord} package available on CRAN.
More specifically, we implement a new multiple measurement object which takes into account the fact that the data contains three dimensions: the subject $i$, the time point $t$ and the outcome $j$. Furthermore, new objects for the error structure among the ordinal responses are implemented to reflect the covariance in Equation~\eqref{eq:corstruct}. After introducing these new objects we exemplify on a simulated data set the use of the estimation function \code{mvordflex()} to estimate the three-dimensional model.  For model comparison, we also show case how three further models can be estimated: one where all correlations
are set to zero, one with cross-sectional correlations but zero longitudinal
correlations and one with longitudinal correlations but zero cross-sectional correlations.

Package \pkg{mvordflex} is available at \url{https://gitlab.com/lauravana/mvordflex} and can be installed using, e.g.,:
\begin{Schunk}
\begin{Sinput}
> library("remotes")
> install_gitlab("lauravana/mvordflex")
> library("mvordflex")
\end{Sinput}
\end{Schunk}
\subsection{Objects implemented to extend \pkg{mvord}}
\subsubsection{Multiple measurement object}
The multiple measurement object \code{MMO3} is implemented, to which the
three dimensional panel data can be passed,  in addition to the existing  multiple measurement object \code{MMO} and
\code{MMO2} in \pkg{mvord}.
The \code{MMO3} object requires the user to
specify the name of the column containing the ordinal observations,
the subject index ($i$), the time index ($t$) and the multiple measurement index
($j$).
Note that all the ordinal observations for all outcomes should be contained in one
column of the data frame to be passed to the model. Moreover, the covariates
are allowed to vary for each $i,t,j$.
The \code{MMO3} object constitutes the left hand side of the formula object.
\begin{Schunk}
\begin{Sinput}
> formula <- MMO3(response, firm_id, year_id, outcome_id) ~ 0 + X1 + ... + Xp
\end{Sinput}
\end{Schunk}
\subsubsection{Error structure objects}
The proposed correlation structure is implemented as a new error structure named
\code{cor\_MMO3()} of class \code{`error\_struct'}.
\begin{Schunk}
\begin{Sinput}
> cor_MMO3(formula = ~1, value = numeric(0), fixed = FALSE, Psi.diag = TRUE)
\end{Sinput}
\end{Schunk}
As discussed in Section~\ref{sec:model}, this correlation structure
consists of $q(q-1)/2$ correlation parameters for the cross-sectional 
structure and $q$ persistence parameters. The argument \code{value} can be
used to specify starting values for the parameters 
$\rho_1, \rho_2, \ldots, \rho_{q(q-1)/2}, \psi_1, \psi_2, \ldots, \psi_q$. 
If argument \code{fixed} is set to \code{TRUE}, the
parameters will be set to \code{value} and not estimated. In the current
implementation the \code{Psi.diag} must be set to \code{TRUE}, as we at the
time of writing do not allow for a general structure on the matrix $\Psi$.
Setting \code{Psi.diag} to \code{FALSE} would estimate a 
general $\Psi$ matrix which satisfies the stationarity constraints of multivariate
autoregressive process. To ensure that the stationarity constraints
are satisfied, a decomposition of $\Psi$ which relies on a positive
definite matrix and an orthogonal matrix as described in \citep{Roy2019} can be employed.
For the positive definite matrix the unconstrained log matrix parameterization
can be employed while for the orthogonal matrix Givens parameterization or the 
Cayley representation can be used. 

We also provide two additional correlation structures, namely
\code{cor\_MMO3\_cross()} and  \code{cor\_MMO3\_ar1()}, which allow the users to
estimate models with only cross-sectional and only longitudinal correlations, respectively.
\begin{Schunk}
\begin{Sinput}
> cor_MMO3_ar1(formula = ~1, value = numeric(0), fixed = FALSE, Psi.diag = TRUE)
\end{Sinput}
\end{Schunk}
\begin{Schunk}
\begin{Sinput}
> cor_MMO3_cross(formula = ~1, value = numeric(0), fixed = FALSE)
\end{Sinput}
\end{Schunk}
The user may note that these two models
can be estimated using \pkg{mvord}. However, if a comparison among different model
based on AIC or BIC is desired, the same pairs should be used in the composition of
the pairwise likelihood in all estimated models to ensure comparability.

\subsubsection{Further objects}
Finally, we provide a new link function \code{mvlogitapprox()}  which uses the 
multivariate Student-$t$ distribution as an approximation to the multivariate
logistic distribution implemented in \code{mvord::mvlogit()} (for more details see 
Section~\ref{sec:mvlink_logit}).

Other functionalities of the \pkg{mvord} package can be
used also for the three dimensional model. Constraints on the threshold parameters and
on the regression coefficients can be set as described in  Section~3.5 and~3.6 of
\cite{pub:mvord:Hirk+Hornik+Vana:2020}, by taking into account that the
dimensionality of the problem is $q \cdot T$ (see structure of correlation
matrix in Equation~\eqref{eq:corstruct}).

\subsection{Simulated data}

We use a simulated data set for software illustration purposes,
given that the data used in Section~\ref{sec:empirical_results} cannot be provided
due to licensing constraints.  The reader should be aware that
the data below is not simulated from the proposed model, so there is no relation
among the covariates and the response and no dependence among the ordinal
responses.
\begin{Schunk}
\begin{Sinput}
> n   <- 100  # number of firms i.e., subjects
> TT  <- 5   # number of time points
> q   <- 3 # number of ordinal responses
> K_R1   <- 7 # number of classes for rater 1
> K_R2   <- 7 # number of classes for rater 2
> K_Fail <- 2 # number of classes for response Fa
> set.seed(1234)
> credit_data <- data.frame(
+   firm_id = rep(1:n, each = TT),
+   year_id = rep(1:TT, n),
+   R1      = sample(1:K_R1,   n * TT, replace = TRUE),
+   R2      = sample(1:K_R2,   n * TT, replace = TRUE),
+   Fail    = sample(1:K_Fail, n * TT, replace = TRUE),
+   X1 = rnorm(n * TT),
+   X2 = rnorm(n * TT),
+   X3 = rnorm(n * TT),
+   X4 = rnorm(n * TT),
+   X5 = rnorm(n * TT),
+   X6 = rnorm(n * TT),
+   X7 = rnorm(n * TT)
+ )
\end{Sinput}
\end{Schunk}
The data frame \code{credit\_data} has the following structure:
\begin{Schunk}
\begin{Sinput}
> head(credit_data)
\end{Sinput}
\begin{Soutput}
  firm_id year_id R1 R2 Fail          X1          X2           X3         X4
1       1       1  4  2    2 -0.05454602 -0.41121274  0.002416472  0.2476992
2       1       2  2  7    2 -0.56820688  0.11820854 -0.717301983  0.1606688
3       1       3  6  2    2 -0.92697594 -0.07116077  0.281338028 -0.1135968
4       1       4  5  5    2 -1.13573518 -0.60869743  0.007281897 -0.7605817
5       1       5  4  6    2  0.91396234  0.32164529 -1.729700226  0.2081472
6       2       1  7  3    2 -0.98606283 -0.91048953 -1.843473616  0.7513504
          X5         X6          X7
1 -0.4946109 -0.4602714  1.88205455
2  1.0714708  0.1580400  3.00836977
3 -1.2049747 -0.4087352  0.61714486
4  1.0217973  1.0342143 -1.13306049
5  1.4212243  0.4099660 -0.41808934
6  0.3853650 -0.6201778 -0.01337428
\end{Soutput}
\end{Schunk}
where for the ratings $1$ represents the worst and $7$ represents the best class
while for the failure indicator a $1$ represents failure and $2$ represents
no failure.

Note that this data set contains the multiple outcomes in the columns and that
the covariates vary only over the subject and time dimensions (as they contain
financial statement information). To bring the data into the format necessary
for the \code{MMO3} object, we make the following manipulations:
\begin{Schunk}
\begin{Sinput}
> df_MMO3 <- rbind(
+   cbind("response" = credit_data$R1,   "outcome_id" = "R1",   credit_data),
+   cbind("response" = credit_data$R2,   "outcome_id" = "R2",   credit_data),
+   cbind("response" = credit_data$Fail, "outcome_id" = "Fail", credit_data))
> head(df_MMO3)
\end{Sinput}
\begin{Soutput}
  response outcome_id firm_id year_id R1 R2 Fail          X1          X2
1        4         R1       1       1  4  2    2 -0.05454602 -0.41121274
2        2         R1       1       2  2  7    2 -0.56820688  0.11820854
3        6         R1       1       3  6  2    2 -0.92697594 -0.07116077
4        5         R1       1       4  5  5    2 -1.13573518 -0.60869743
5        4         R1       1       5  4  6    2  0.91396234  0.32164529
6        7         R1       2       1  7  3    2 -0.98606283 -0.91048953
            X3         X4         X5         X6          X7
1  0.002416472  0.2476992 -0.4946109 -0.4602714  1.88205455
2 -0.717301983  0.1606688  1.0714708  0.1580400  3.00836977
3  0.281338028 -0.1135968 -1.2049747 -0.4087352  0.61714486
4  0.007281897 -0.7605817  1.0217973  1.0342143 -1.13306049
5 -1.729700226  0.2081472  1.4212243  0.4099660 -0.41808934
6 -1.843473616  0.7513504  0.3853650 -0.6201778 -0.01337428
\end{Soutput}
\end{Schunk}
\subsection{Estimation of the models on simulated data}
\subsubsection{Three-dimensional model}
To estimate the proposed model with different sets of coefficients and thresholds
for each outcome using the logit link the following code can be used. In order to improve
computational speed, we only consider pairs of observations in the composite likelihood
which are at most one time point apart (by setting \code{PL.lag = 1}). Also,
we specify constraints on the threshold and regression coefficients such that
they are equal across time but varying across outcomes.
\begin{Schunk}
\begin{Sinput}
> pl.lag <- 1
> res_logit <- mvordflex(
+   MMO3(response, firm_id, year_id, outcome_id) ~ 0 + X1 + X2 + X3 + X4 + X5 + X6 + X7,
+   data = df_MMO3,
+   link = mvord::mvlogit(),
+   error.structure = cor_MMO3(~1),
+   coef.constraints = rep(1:q,TT),
+   threshold.constraints = rep(1:q,TT),
+   PL.lag = pl.lag)
\end{Sinput}
\end{Schunk}
The results of the model can be inspected using the \code{summary} function:
\begin{Schunk}
\begin{Sinput}
> summary(res_logit)
\end{Sinput}
\begin{Soutput}
Call: mvordflex(formula = MMO3(response, firm_id, year_id, outcome_id) ~ 
    0 + X1 + X2 + X3 + X4 + X5 + X6 + X7, data = df_MMO3, error.structure = cor_MMO3(~1), 
    link = mvord::mvlogit(), coef.constraints = rep(1:q, TT), 
    threshold.constraints = rep(1:q, TT), PL.lag = pl.lag)

Formula: MMO3(response, firm_id, year_id, outcome_id) ~ 0 + X1 + X2 + 
    X3 + X4 + X5 + X6 + X7

   link threshold nsubjects ndim     logPL    CLAIC    CLBIC fevals
mvlogit  flexible       100   15 -15423.96 31655.33 32707.06   3177

Thresholds:
          Estimate Std. Error  z value  Pr(>|z|)    
1R1 1|2   -2.14126    0.19879 -10.7716 < 2.2e-16 ***
1R1 2|3   -1.21920    0.13826  -8.8182 < 2.2e-16 ***
1R1 3|4   -0.42391    0.12360  -3.4297 0.0006043 ***
1R1 4|5    0.15929    0.12045   1.3225 0.1859975    
1R1 5|6    0.74591    0.13410   5.5623 2.662e-08 ***
1R1 6|7    1.76528    0.16816  10.4978 < 2.2e-16 ***
1R2 1|2   -1.63667    0.14525 -11.2683 < 2.2e-16 ***
1R2 2|3   -0.90287    0.13122  -6.8803 5.971e-12 ***
1R2 3|4   -0.21442    0.12460  -1.7208 0.0852878 .  
1R2 4|5    0.22619    0.12100   1.8693 0.0615856 .  
1R2 5|6    0.78534    0.12811   6.1303 8.773e-10 ***
1R2 6|7    1.64813    0.15237  10.8169 < 2.2e-16 ***
1Fail 1|2 -0.15125    0.11389  -1.3281 0.1841508    
---
Signif. codes:  0 ‘***’ 0.001 ‘**’ 0.01 ‘*’ 0.05 ‘.’ 0.1 ‘ ’ 1

Coefficients:
        Estimate  Std. Error z value Pr(>|z|)  
X1 1  0.01624812  0.10404012  0.1562  0.87590  
X1 2  0.02907773  0.10255632  0.2835  0.77677  
X1 3 -0.17356886  0.12162026 -1.4271  0.15354  
X2 1  0.01881268  0.09879546  0.1904  0.84898  
X2 2 -0.09315172  0.10486838 -0.8883  0.37439  
X2 3  0.04518789  0.11296786  0.4000  0.68915  
X3 1  0.09568357  0.10008363  0.9560  0.33905  
X3 2  0.00059016  0.10669668  0.0055  0.99559  
X3 3 -0.09215283  0.10787125 -0.8543  0.39295  
X4 1  0.00058229  0.10514732  0.0055  0.99558  
X4 2 -0.09203892  0.10910576 -0.8436  0.39891  
X4 3 -0.12544095  0.12431014 -1.0091  0.31293  
X5 1 -0.09291894  0.10380294 -0.8951  0.37071  
X5 2  0.15000511  0.09707369  1.5453  0.12228  
X5 3  0.00873242  0.12280489  0.0711  0.94331  
X6 1  0.10317883  0.10189018  1.0126  0.31123  
X6 2  0.01551424  0.10205154  0.1520  0.87917  
X6 3 -0.02293986  0.12291411 -0.1866  0.85195  
X7 1 -0.17414693  0.10533221 -1.6533  0.09827 .
X7 2 -0.04256653  0.11245883 -0.3785  0.70505  
X7 3 -0.05434595  0.11974357 -0.4539  0.64993  
---
Signif. codes:  0 ‘***’ 0.001 ‘**’ 0.01 ‘*’ 0.05 ‘.’ 0.1 ‘ ’ 1

Error Structure:
          Estimate Std. Error z value Pr(>|z|)   
corr 1 2  0.106424   0.062828  1.6939  0.09029 . 
corr 1 3  0.039619   0.070658  0.5607  0.57499   
corr 2 3  0.035688   0.069927  0.5104  0.60980   
psi 1     0.083848   0.078342  1.0703  0.28449   
psi 2    -0.199654   0.071432 -2.7950  0.00519 **
psi 3     0.119924   0.101317  1.1837  0.23655   
---
Signif. codes:  0 ‘***’ 0.001 ‘**’ 0.01 ‘*’ 0.05 ‘.’ 0.1 ‘ ’ 1
\end{Soutput}
\end{Schunk}
We observe that the results are in line to what we would expect:
the regression coefficients are insignificant as well as all
persistence parameter at a 5\% significance level.

\subsubsection{Model with  cross-sectional correlation structure}
For model comparison, we estimate a model with cross-sectional correlation among the responses using
the \pkg{mvordflex} package:
\begin{Schunk}
\begin{Sinput}
> library("mvordflex")
> res_cross_logit <- mvordflex(
+   MMO3(response, firm_id, year_id, outcome_id) ~ 0 + X1 + X2 + X3 + X4 + X5 + X6 + X7,
+   data = df_MMO3,
+   link = mvord::mvlogit(),
+   error.structure = cor_MMO3_cross(~1),
+   coef.constraints = rep(1:q, TT),
+   threshold.constraints = rep(1:q, TT),
+   PL.lag = pl.lag)
\end{Sinput}
\end{Schunk}
Note that this model can also be constructed using the \code{cor\_general()}
structure in \pkg{mvord} (see code below). However, given that the  \pkg{mvord}
model would contain different subject units (i.e., firm-years vs. firms), a
comparison using CLAIC or CLBIC would not be possible. The models would then
need to be compared based on their prediction ability in- or out-of-sample.
\begin{Schunk}
\begin{Sinput}
> library("mvord")
> df_MMO3$firm_year_id <-
+   paste(df_MMO3$firm_id, df_MMO3$year_id, sep = "+")
> res_cross_logit_mvord <- mvord::mvord(
+   MMO(response, firm_year_id, outcome_id) ~ 0 + X1 + X2 + X3 + X4 + X5 + X6 + X7,
+   data = df_MMO3,
+   link = mvord::mvlogit(),
+   error.structure =  cor_general())
\end{Sinput}
\end{Schunk}
\subsubsection{Model with longitudinal correlation structure}
The estimation of the model with longitudinal but no cross-sectional correlation can be performed using:
\begin{Schunk}
\begin{Sinput}
> library("mvordflex")
> res_ar1_logit <- mvordflex(
+   MMO3(response, firm_id, year_id, outcome_id) ~ 0 + X1 + X2 + X3 + X4 + X5 + X6 + X7,
+   data = df_MMO3,
+   link = mvord::mvlogit(),
+   error.structure = cor_MMO3_ar1(~1),
+   coef.constraints = rep(1:q, TT),
+   threshold.constraints = rep(1:q, TT),
+   PL.lag = pl.lag)
\end{Sinput}
\end{Schunk}
This model can be replicated by three different models constructed using the
\code{cor\_ar1()} structure in \pkg{mvord}, one for each outcome. Note, again, that
these models cannot be directly compared in terms of information criteria with the
model from \pkg{mvordflex}.

\subsubsection{Model with no correlation among the responses}
Finally, we would like to compare the models estimated above with a baseline model, i.e., a model with no correlation among the responses can be estimated using
the \pkg{mvord} package. For this purpose we create a new column in
the data frame \code{df\_MMO3} which contains the combined year and outcome ID
and which will serve as the multiple measurement index.
\begin{Schunk}
\begin{Sinput}
> df_MMO3$year_outcome_id <- factor(
+   paste(df_MMO3$year_id, df_MMO3$outcome_id, sep = "+"))
> levels(df_MMO3$year_outcome_id)
\end{Sinput}
\begin{Soutput}
 [1] "1+Fail" "1+R1"   "1+R2"   "2+Fail" "2+R1"   "2+R2"   "3+Fail" "3+R1"  
 [9] "3+R2"   "4+Fail" "4+R1"   "4+R2"   "5+Fail" "5+R1"   "5+R2"  
\end{Soutput}
\end{Schunk}
Given that we only want to consider pairs of observations
which are at most one time point apart (as in the model above),
we can specify in the \code{control} argument of function \code{mvord()} which
combinations of responses should enter the pairwise likelihood in the form of
a list. For the 15 responses in this example, there are in total 105 pairs of
responses.
\begin{Schunk}
\begin{Sinput}
> v <- seq_len(nlevels(df_MMO3$year_outcome_id))
> names(v) <- levels(df_MMO3$year_outcome_id)
> combis_all <- combn(v, 2, simplify = FALSE)
> head(combis_all)
\end{Sinput}
\begin{Soutput}
[[1]]
1+Fail   1+R1 
     1      2 

[[2]]
1+Fail   1+R2 
     1      3 

[[3]]
1+Fail 2+Fail 
     1      4 

[[4]]
1+Fail   2+R1 
     1      5 

[[5]]
1+Fail   2+R2 
     1      6 

[[6]]
1+Fail 3+Fail 
     1      7 
\end{Soutput}
\end{Schunk}
However, we only consider a subset:
\begin{Schunk}
\begin{Sinput}
> id_keep <- sapply(combis_all, function(x) {
+   abs(diff(match(gsub("\\+.*", "", names(x)), unique(df_MMO3$year_id)))) <= pl.lag
+ })
> combis <- combis_all[id_keep]
> length(combis)
\end{Sinput}
\begin{Soutput}
[1] 51
\end{Soutput}
\end{Schunk}
We see that from the 105 combinations only 51 will be used.
\begin{Schunk}
\begin{Sinput}
> library("mvord")
> res_ident_logit <- mvord(
+   MMO(response, firm_id, year_outcome_id) ~ 0 + X1 + X2 + X3 + X4 + X5 + X6 + X7,
+   data = df_MMO3,
+   link = mvord::mvlogit(),
+   error.structure = cor_equi(~ 1, value = 0, fixed = TRUE),
+   coef.constraints = rep(1:q, TT),
+   threshold.constraints = rep(1:q, TT),
+   control = mvord:: mvord.control(se = TRUE, combis = combis))
\end{Sinput}
\end{Schunk}
\subsubsection{Model comparison}
We can compare all models using the \code{AIC()} function:
\begin{Schunk}
\begin{Sinput}
> AIC(res_ident_logit, res_cross_logit, res_ar1_logit, res_logit)
\end{Sinput}
\begin{Soutput}
                      df      AIC
res_ident_logit 359.3323 31572.39
res_cross_logit 380.2915 31612.39
res_ar1_logit   380.7368 31611.51
res_logit       403.7086 31655.33
\end{Soutput}
\begin{Sinput}
> BIC(res_ident_logit, res_cross_logit, res_ar1_logit, res_logit)
\end{Sinput}
\begin{Soutput}
                      df      BIC
res_ident_logit 359.3323 32508.51
res_cross_logit 380.2915 32603.12
res_ar1_logit   380.7368 32603.40
res_logit       403.7086 32707.06
\end{Soutput}
\end{Schunk}
As expected from  simulation of the data above,
the model with an identity correlation matrix performs best in terms of CLAIC and CLBIC.

\section{Conclusion}\label{sec:concl}
We propose a multivariate ordinal regression model which accounts for
dependence between  repeated and multiple ordinal measurements.
This is achieved by imposing a multivariate autoregressive structure on the
errors underlying the ordinal responses, where the contemporaneous errors
have a general correlation structure and the coefficients of the AR(1) process
capture persistence of the ordinal outcomes over time.
The estimation is performed using composite likelihood methods and a
simulation study confirms that the model
parameters can be recovered well and that the pairwise likelihood approach is 
competitive when compared to a Bayesian approach, both in terms of 
computation time and accuracy of the estimates.
Furthermore, we present the implementation of the model as an \proglang{R}
\pkg{mvordflex}, which is an extension to the \proglang{R}
package \pkg{mvord} and exemplify how users can use the functionality provided
by this extension. 

Finally, we illustrate the framework on a data set containing default and credit
rating information from S\&P and Moody's for US listed companies over the
period 2003--2013. We find that
the proposed model improves the fit when compared to simpler specifications which
take only the cross-sectional correlations or only the time dependence into account.

One of the limitations of the model, which relates mainly to the pairwise likelihood estimation 
is the possibility of overoptimistic standard errors for the threshold parameters. This can be seen from the 
provided simulation studies and has also been documented 
before in the literature \citep[see e.g.,][]{Varin09}. 
Employing re-sampling techniques such as the
jackknife or bootstrap can alleviate this problem and can lead to less biased
variance estimates. However, it comes at a significant computational cost. 

Finally, in our modeling approach we assumed that the observations are 
missing completely at random. In longitudinal studies, this is often not the case,
so approaches which can model the missing data mechanism jointly with the 
observations should be investigated in future research for the panel data case
\citep[see e.g.,][for approaches for longitudinal models]{li2013pairwise, LI201925}.
We also observe that in the presence of highly imbalanced binary responses, 
the sign of some of the error structure parameters is computationally 
unidentifiable and the estimation procedure is inaccurate. Up-sampling can alleviate this problem. Alternatively, the model can further be extended to accommodate 
for different multivariate (asymmetric) link functions. This could prove beneficial for the modeling of imbalanced responses.
A further extension would be modeling $\Psi$
as a full matrix, with certain constraints to ensure stationarity and 
invertibility of the multivariate AR(1) process. Such a 
specification can prove relevant in e.g.,~economic applications. 

\section*{Computational details}
The package \pkg{mvordflex} is provided at \url{https://gitlab.com/lauravana/mvordflex}.
The codes for reproducing the three simulation exercises as well as the
the tables and figures from Section~\ref{sec:simulation}
can also be found at \url{https://gitlab.com/lauravana/mvordflex/paper}.
We provide an implementation of multivariate ordinal regression models with probit link
for two-dimensional ordinal panel data in the following
repository: \url{https://github.com/lauravana/mvordstanr}.

\section*{Acknowledgments}
This research was supported by funds of the Oesterreichische Nationalbank (Austrian
Central Bank, Anniversary Fund, project number: 18482 ``Multivariate ordinal regression models for
enhanced credit risk modeling'').
The author would like to thank Rainer Hirk for valuable input on an earlier version 
of the manuscript, as well as the two anonymous referees and the associate editor 
for their insightful feedback.

\bibliographystyle{elsarticle-harv}
\bibliography{refs}

\begin{thebibliography}{45}
\expandafter\ifx\csname natexlab\endcsname\relax\def\natexlab#1{#1}\fi
\providecommand{\url}[1]{\texttt{#1}}
\providecommand{\href}[2]{#2}
\providecommand{\path}[1]{#1}
\providecommand{\DOIprefix}{doi:}
\providecommand{\ArXivprefix}{arXiv:}
\providecommand{\URLprefix}{URL: }
\providecommand{\Pubmedprefix}{pmid:}
\providecommand{\doi}[1]{\href{http://dx.doi.org/#1}{\path{#1}}}
\providecommand{\Pubmed}[1]{\href{pmid:#1}{\path{#1}}}
\providecommand{\bibinfo}[2]{#2}
\ifx\xfnm\relax \def\xfnm[#1]{\unskip,\space#1}\fi
\bibitem[{Alpuim and El-Shaarawi(2008)}]{alpuim2008efficiency}
\bibinfo{author}{Alpuim, T.}, \bibinfo{author}{El-Shaarawi, A.},
  \bibinfo{year}{2008}.
\newblock \bibinfo{title}{On the efficiency of regression analysis with ar (p)
  errors}.
\newblock \bibinfo{journal}{Journal of Applied Statistics}
  \bibinfo{volume}{35}, \bibinfo{pages}{717--737}.
\bibitem[{Anderson(1954)}]{anderson1954problem}
\bibinfo{author}{Anderson, R.L.}, \bibinfo{year}{1954}.
\newblock \bibinfo{title}{The problem of autocorrelation in regression
  analysis}.
\newblock \bibinfo{journal}{Journal of the American Statistical Association}
  \bibinfo{volume}{49}, \bibinfo{pages}{113--129}.
\bibitem[{{Bank of International Settlements}(2004)}]{baselii}
\bibinfo{author}{{Bank of International Settlements}}, \bibinfo{year}{2004}.
\newblock \bibinfo{title}{Basel {II}: International convergence of capital
  measurement and capital standards: A revised framework}.
\newblock \bibinfo{howpublished}{Online publication}.
\newblock \URLprefix \url{http://www.bis.org/publ/bcbs107.htm}.
\bibitem[{Bartolucci and Farcomeni(2009)}]{bartolucci2009multivariate}
\bibinfo{author}{Bartolucci, F.}, \bibinfo{author}{Farcomeni, A.},
  \bibinfo{year}{2009}.
\newblock \bibinfo{title}{A multivariate extension of the dynamic logit model
  for longitudinal data based on a latent {M}arkov heterogeneity structure}.
\newblock \bibinfo{journal}{Journal of the American Statistical Association}
  \bibinfo{volume}{104}, \bibinfo{pages}{816--831}.
\newblock \DOIprefix\doi{10.1198/jasa.2009.0107}.
\bibitem[{Betancourt and Girolami(2015)}]{betancourt2015hamiltonian}
\bibinfo{author}{Betancourt, M.}, \bibinfo{author}{Girolami, M.},
  \bibinfo{year}{2015}.
\newblock \bibinfo{title}{Hamiltonian monte carlo for hierarchical models}.
\newblock \bibinfo{journal}{Current trends in Bayesian methodology with
  applications} \bibinfo{volume}{79}, \bibinfo{pages}{2--4}.
\bibitem[{Bhat et~al.(2010)Bhat, Varin and Ferdous}]{bhat2010comparison}
\bibinfo{author}{Bhat, C.R.}, \bibinfo{author}{Varin, C.},
  \bibinfo{author}{Ferdous, N.}, \bibinfo{year}{2010}.
\newblock \bibinfo{title}{A comparison of the maximum simulated likelihood and
  composite marginal likelihood estimation approaches in the context of the
  multivariate ordered-response model}, in: \bibinfo{booktitle}{Maximum
  simulated likelihood methods and applications}. \bibinfo{publisher}{Emerald
  Group Publishing Limited}.
\newblock \DOIprefix\doi{10.1108/S0731-9053(2010)0000026007}.
\bibitem[{Cagnone et~al.(2009)Cagnone, Moustaki and
  Vasdekis}]{cagnone2009latent}
\bibinfo{author}{Cagnone, S.}, \bibinfo{author}{Moustaki, I.},
  \bibinfo{author}{Vasdekis, V.}, \bibinfo{year}{2009}.
\newblock \bibinfo{title}{Latent variable models for multivariate longitudinal
  ordinal responses}.
\newblock \bibinfo{journal}{British Journal of Mathematical and Statistical
  Psychology} \bibinfo{volume}{62}, \bibinfo{pages}{401--415}.
\newblock \DOIprefix\doi{10.1348/000711008X320134}.
\bibitem[{Campbell et~al.(2008)Campbell, Hilscher and
  Szilagyi}]{campbell2008search}
\bibinfo{author}{Campbell, J.Y.}, \bibinfo{author}{Hilscher, J.},
  \bibinfo{author}{Szilagyi, J.}, \bibinfo{year}{2008}.
\newblock \bibinfo{title}{In search of distress risk}.
\newblock \bibinfo{journal}{The Journal of Finance} \bibinfo{volume}{63},
  \bibinfo{pages}{2899--2939}.
\newblock \DOIprefix\doi{10.3386/w12362}.
\bibitem[{Chaubert et~al.(2008)Chaubert, Mortier and
  Saint~Andr{\'e}}]{chaubert2008multivariate}
\bibinfo{author}{Chaubert, F.}, \bibinfo{author}{Mortier, F.},
  \bibinfo{author}{Saint~Andr{\'e}, L.}, \bibinfo{year}{2008}.
\newblock \bibinfo{title}{Multivariate dynamic model for ordinal outcomes}.
\newblock \bibinfo{journal}{Journal of Multivariate Analysis}
  \bibinfo{volume}{99}, \bibinfo{pages}{1717--1732}.
\newblock \DOIprefix\doi{10.1016/j.jmva.2008.01.011}.
\bibitem[{Chib(1993)}]{chib1993bayes}
\bibinfo{author}{Chib, S.}, \bibinfo{year}{1993}.
\newblock \bibinfo{title}{Bayes regression with autoregressive errors: A
  {G}ibbs sampling approach}.
\newblock \bibinfo{journal}{Journal of Econometrics} \bibinfo{volume}{58},
  \bibinfo{pages}{275--294}.
\newblock \DOIprefix\doi{10.1016/0304-4076(93)90046-8}.
\bibitem[{Christensen(2023)}]{pkgordinal}
\bibinfo{author}{Christensen, R.H.B.}, \bibinfo{year}{2023}.
\newblock \bibinfo{title}{ordinal---Regression Models for Ordinal Data}.
\newblock \URLprefix \url{https://CRAN.R-project.org/package=ordinal}.
  \bibinfo{note}{r package version 2023.12-4}.
\bibitem[{Cochrane and Orcutt(1949)}]{cochrane1949application}
\bibinfo{author}{Cochrane, D.}, \bibinfo{author}{Orcutt, G.H.},
  \bibinfo{year}{1949}.
\newblock \bibinfo{title}{Application of least squares regression to
  relationships containing auto-correlated error terms}.
\newblock \bibinfo{journal}{Journal of the American statistical association}
  \bibinfo{volume}{44}, \bibinfo{pages}{32--61}.
\newblock \DOIprefix\doi{10.1080/01621459.1954.10501219}.
\bibitem[{Durbin(1960)}]{durbin1960estimation}
\bibinfo{author}{Durbin, J.}, \bibinfo{year}{1960}.
\newblock \bibinfo{title}{Estimation of parameters in time-series regression
  models}.
\newblock \bibinfo{journal}{Journal of the royal statistical society: Series B
  (Methodological)} \bibinfo{volume}{22}, \bibinfo{pages}{139--153}.
\newblock \DOIprefix\doi{10.1111/j.2517-6161.1960.tb00361.x}.
\bibitem[{Ferrari et~al.(2016)Ferrari, Qian and
  Hunter}]{ferrari2016parsimonious}
\bibinfo{author}{Ferrari, D.}, \bibinfo{author}{Qian, G.},
  \bibinfo{author}{Hunter, T.}, \bibinfo{year}{2016}.
\newblock \bibinfo{title}{Parsimonious and efficient likelihood composition by
  gibbs sampling}.
\newblock \bibinfo{journal}{Journal of Computational and Graphical Statistics}
  \bibinfo{volume}{25}, \bibinfo{pages}{935--953}.
\newblock \DOIprefix\doi{10.1080/10618600.2015.1058799}.
\bibitem[{Greenwell et~al.(2017)Greenwell, McCarthy and Boehmke}]{pkg:sure}
\bibinfo{author}{Greenwell, B.}, \bibinfo{author}{McCarthy, A.},
  \bibinfo{author}{Boehmke, B.}, \bibinfo{year}{2017}.
\newblock \bibinfo{title}{sure: Surrogate Residuals for Ordinal and General
  Regression Models}.
\newblock \URLprefix \url{https://CRAN.R-project.org/package=sure}.
  \bibinfo{note}{r package version 0.2.0}.
\bibitem[{Guney et~al.(2022)Guney, Arslan and Yavuz}]{guney2022robust}
\bibinfo{author}{Guney, Y.}, \bibinfo{author}{Arslan, O.},
  \bibinfo{author}{Yavuz, F.G.}, \bibinfo{year}{2022}.
\newblock \bibinfo{title}{Robust estimation in multivariate heteroscedastic
  regression models with autoregressive covariance structures using em
  algorithm}.
\newblock \bibinfo{journal}{Journal of Multivariate Analysis}
  \bibinfo{volume}{191}, \bibinfo{pages}{105026}.
\bibitem[{Hirk et~al.(2019)Hirk, Hornik and Vana}]{pub:Hirk+Hornik+Vana:2018a}
\bibinfo{author}{Hirk, R.}, \bibinfo{author}{Hornik, K.},
  \bibinfo{author}{Vana, L.}, \bibinfo{year}{2019}.
\newblock \bibinfo{title}{Multivariate ordinal regression models: An analysis
  of corporate credit ratings}.
\newblock \bibinfo{journal}{Statistical Methods {\&} Applications} ,
  \bibinfo{pages}{507--539}\DOIprefix\doi{10.1007/s10260-018-00437-7}.
\bibitem[{Hirk et~al.(2020)Hirk, Hornik and
  Vana}]{pub:mvord:Hirk+Hornik+Vana:2020}
\bibinfo{author}{Hirk, R.}, \bibinfo{author}{Hornik, K.},
  \bibinfo{author}{Vana, L.}, \bibinfo{year}{2020}.
\newblock \bibinfo{title}{\pkg{mvord}: An \proglang{R} package for fitting
  multivariate ordinal regression models}.
\newblock \bibinfo{journal}{Journal of Statistical Software}
  \bibinfo{volume}{93}, \bibinfo{pages}{1--41}.
\newblock \DOIprefix\doi{10.18637/jss.v093.i04}.
\bibitem[{Hirk and Vana(2024)}]{pkgmvordflex}
\bibinfo{author}{Hirk, R.}, \bibinfo{author}{Vana, L.}, \bibinfo{year}{2024}.
\newblock \bibinfo{title}{{mvordflex}: Multivariate Ordinal Regression Models
  -- Extension to Three Dimensional Panel Data}.
\newblock \URLprefix \url{https://gitlab.com/lauravana/mvordflex}.
  \bibinfo{note}{r package version 0.0.2}.
\bibitem[{Hirk et~al.(2022)Hirk, Vana and Hornik}]{HIRK2022224}
\bibinfo{author}{Hirk, R.}, \bibinfo{author}{Vana, L.},
  \bibinfo{author}{Hornik, K.}, \bibinfo{year}{2022}.
\newblock \bibinfo{title}{A corporate credit rating model with autoregressive
  errors}.
\newblock \bibinfo{journal}{Journal of Empirical Finance} \bibinfo{volume}{69},
  \bibinfo{pages}{224--240}.
\newblock \DOIprefix\doi{https://doi.org/10.1016/j.jempfin.2022.09.002}.
\bibitem[{Hirk et~al.(2021)Hirk, Vana, Pichler and Hornik}]{Hirk2021jcr}
\bibinfo{author}{Hirk, R.}, \bibinfo{author}{Vana, L.},
  \bibinfo{author}{Pichler, S.}, \bibinfo{author}{Hornik, K.},
  \bibinfo{year}{2021}.
\newblock \bibinfo{title}{A joint model of failures and credit ratings}.
\newblock \bibinfo{journal}{Journal of Credit Risk} \bibinfo{volume}{17},
  \bibinfo{pages}{61--88}.
\newblock \DOIprefix\doi{10.21314/JCR.2020.264}.
\bibitem[{{John C. Nash}(2014)}]{jssoptimx}
\bibinfo{author}{{John C. Nash}}, \bibinfo{year}{2014}.
\newblock \bibinfo{title}{On best practice optimization methods in {R}}.
\newblock \bibinfo{journal}{Journal of Statistical Software}
  \bibinfo{volume}{60}, \bibinfo{pages}{1--14}.
\newblock \DOIprefix\doi{10.18637/jss.v060.i02}.
\bibitem[{Kenne~Pagui and Canale(2016)}]{Pagui2015}
\bibinfo{author}{Kenne~Pagui, E.C.}, \bibinfo{author}{Canale, A.},
  \bibinfo{year}{2016}.
\newblock \bibinfo{title}{Pairwise likelihood inference for multivariate
  ordinal responses with applications to customer satisfaction}.
\newblock \bibinfo{journal}{Applied Stochastic Models in Business and Industry}
  \bibinfo{volume}{32}, \bibinfo{pages}{273--282}.
\newblock \DOIprefix\doi{10.1002/asmb.2147}.
\bibitem[{Li and Grace(2013)}]{li2013pairwise}
\bibinfo{author}{Li, H.}, \bibinfo{author}{Grace, Y.Y.}, \bibinfo{year}{2013}.
\newblock \bibinfo{title}{A pairwise likelihood approach for longitudinal data
  with missing observations in both response and covariates}.
\newblock \bibinfo{journal}{Computational Statistics \& Data Analysis}
  \bibinfo{volume}{68}, \bibinfo{pages}{66--81}.
\newblock \DOIprefix\doi{10.1016/j.csda.2013.06.001}.
\bibitem[{Li et~al.(2019)Li, Shu, He and Yi}]{LI201925}
\bibinfo{author}{Li, H.}, \bibinfo{author}{Shu, D.}, \bibinfo{author}{He, W.},
  \bibinfo{author}{Yi, G.Y.}, \bibinfo{year}{2019}.
\newblock \bibinfo{title}{Variable selection via the composite likelihood
  method for multilevel longitudinal data with missing responses and
  covariates}.
\newblock \bibinfo{journal}{Computational Statistics \& Data Analysis}
  \bibinfo{volume}{135}, \bibinfo{pages}{25--34}.
\newblock \DOIprefix\doi{10.1016/j.csda.2019.01.011}.
\bibitem[{Lin et~al.(2021)Lin, Mermelstein and Hedeker}]{Lin2021}
\bibinfo{author}{Lin, X.}, \bibinfo{author}{Mermelstein, R.},
  \bibinfo{author}{Hedeker, D.}, \bibinfo{year}{2021}.
\newblock \bibinfo{title}{Analysis of multivariate longitudinal substance use
  outcomes using multivariate mixed cumulative logit model}.
\newblock \bibinfo{journal}{BMC Medical Research Methodology}
  \bibinfo{volume}{21}, \bibinfo{pages}{239}.
\newblock \URLprefix \url{https://doi.org/10.1186/s12874-021-01444-1},
  \DOIprefix\doi{10.1186/s12874-021-01444-1}.
\bibitem[{Liu and Zhang(2018)}]{doi:10.1080/01621459.2017.1292915}
\bibinfo{author}{Liu, D.}, \bibinfo{author}{Zhang, H.}, \bibinfo{year}{2018}.
\newblock \bibinfo{title}{Residuals and diagnostics for ordinal regression
  models: A surrogate approach}.
\newblock \bibinfo{journal}{Journal of the American Statistical Association}
  \bibinfo{volume}{113}, \bibinfo{pages}{845--854}.
\newblock \URLprefix \url{https://doi.org/10.1080/01621459.2017.1292915},
  \DOIprefix\doi{10.1080/01621459.2017.1292915},
  \href{http://arxiv.org/abs/https://doi.org/10.1080/01621459.2017.1292915}{{\tt
  arXiv:https://doi.org/10.1080/01621459.2017.1292915}}. \bibinfo{note}{pMID:
  30220754}.
\bibitem[{Liu and Hedeker(2006)}]{liu2006mixed}
\bibinfo{author}{Liu, L.C.}, \bibinfo{author}{Hedeker, D.},
  \bibinfo{year}{2006}.
\newblock \bibinfo{title}{A mixed-effects regression model for longitudinal
  multivariate ordinal data}.
\newblock \bibinfo{journal}{Biometrics} \bibinfo{volume}{62},
  \bibinfo{pages}{261--268}.
\bibitem[{O'{B}rien and Dunson(2004)}]{OBrien2004}
\bibinfo{author}{O'{B}rien, S.M.}, \bibinfo{author}{Dunson, D.B.},
  \bibinfo{year}{2004}.
\newblock \bibinfo{title}{Bayesian multivariate logistic regression}.
\newblock \bibinfo{journal}{Biometrics} \bibinfo{volume}{60},
  \bibinfo{pages}{739--746}.
\newblock \DOIprefix\doi{10.1111/j.0006-341X.2004.00224.x}.
\bibitem[{Reusens and Croux(2017)}]{Reusens2017}
\bibinfo{author}{Reusens, P.}, \bibinfo{author}{Croux, C.},
  \bibinfo{year}{2017}.
\newblock \bibinfo{title}{Sovereign credit rating determinants: A comparison
  before and after the {E}uropean debt crisis}.
\newblock \bibinfo{journal}{Journal of Banking \& Finance}
  \bibinfo{volume}{77}, \bibinfo{pages}{108--121}.
\newblock \DOIprefix\doi{10.1016/j.jbankfin.2017.01.006}.
\bibitem[{Roy et~al.(2019)Roy, Mcelroy and Linton}]{Roy2019}
\bibinfo{author}{Roy, A.}, \bibinfo{author}{Mcelroy, T.S.},
  \bibinfo{author}{Linton, P.}, \bibinfo{year}{2019}.
\newblock \bibinfo{title}{Constrained estimation of causal invertible varma}.
\newblock \bibinfo{journal}{Statistica Sinica} \bibinfo{volume}{29},
  \bibinfo{pages}{455--478}.
\newblock \URLprefix \url{https://www.jstor.org/stable/26563263}.
\bibitem[{Schliep et~al.(2021)Schliep, Schafer and
  Hawkey}]{SchliepSchaferHawkey+2021+241+254}
\bibinfo{author}{Schliep, E.M.}, \bibinfo{author}{Schafer, T.L.J.},
  \bibinfo{author}{Hawkey, M.}, \bibinfo{year}{2021}.
\newblock \bibinfo{title}{Distributed lag models to identify the cumulative
  effects of training and recovery in athletes using multivariate ordinal
  wellness data}.
\newblock \bibinfo{journal}{Journal of Quantitative Analysis in Sports}
  \bibinfo{volume}{17}, \bibinfo{pages}{241--254}.
\newblock \DOIprefix\doi{doi:10.1515/jqas-2020-0051}.
\bibitem[{Scott and Kanaroglou(2002)}]{Scott02}
\bibinfo{author}{Scott, D.M.}, \bibinfo{author}{Kanaroglou, P.S.},
  \bibinfo{year}{2002}.
\newblock \bibinfo{title}{An activity-episode generation model that captures
  interactions between household heads: development and empirical analysis}.
\newblock \bibinfo{journal}{Transportation Research Part B: Methodological}
  \bibinfo{volume}{36}, \bibinfo{pages}{875--896}.
\newblock \DOIprefix\doi{10.1016/S0191-2615(01)00039-X}.
\bibitem[{{Stan Development Team}(2024)}]{pkg:rstan}
\bibinfo{author}{{Stan Development Team}}, \bibinfo{year}{2024}.
\newblock \bibinfo{title}{{RStan}: the {R} interface to {Stan}}.
\newblock \URLprefix \url{https://mc-stan.org/}. \bibinfo{note}{r package
  version 2.32.5}.
\bibitem[{Tian et~al.(2015)Tian, Yu and Guo}]{Tian2015}
\bibinfo{author}{Tian, S.}, \bibinfo{author}{Yu, Y.}, \bibinfo{author}{Guo,
  H.}, \bibinfo{year}{2015}.
\newblock \bibinfo{title}{Variable selection and corporate bankruptcy
  forecasts}.
\newblock \bibinfo{journal}{Journal of Banking \& Finance}
  \bibinfo{volume}{52}, \bibinfo{pages}{89--100}.
\newblock \DOIprefix\doi{10.1016/j.jbankfin.2014.12.003}.
\bibitem[{Tua{\c{c}} et~al.(2020)Tua{\c{c}}, G{\"u}ney and
  Arslan}]{tuacc2020parameter}
\bibinfo{author}{Tua{\c{c}}, Y.}, \bibinfo{author}{G{\"u}ney, Y.},
  \bibinfo{author}{Arslan, O.}, \bibinfo{year}{2020}.
\newblock \bibinfo{title}{Parameter estimation of regression model with ar (p)
  error terms based on skew distributions with em algorithm}.
\newblock \bibinfo{journal}{Soft Computing} \bibinfo{volume}{24},
  \bibinfo{pages}{3309--3330}.
\newblock \DOIprefix\doi{10.1007/s00500-019-04089-x}.
\bibitem[{Tua{\c{c}} et~al.(2018)Tua{\c{c}}, G{\"u}ney, {\c{S}}eno{\u{g}}lu and
  Arslan}]{tuacc2018robust}
\bibinfo{author}{Tua{\c{c}}, Y.}, \bibinfo{author}{G{\"u}ney, Y.},
  \bibinfo{author}{{\c{S}}eno{\u{g}}lu, B.}, \bibinfo{author}{Arslan, O.},
  \bibinfo{year}{2018}.
\newblock \bibinfo{title}{Robust parameter estimation of regression model with
  ar (p) error terms}.
\newblock \bibinfo{journal}{Communications in Statistics-Simulation and
  Computation} \bibinfo{volume}{47}, \bibinfo{pages}{2343--2359}.
\bibitem[{Tuzcuoglu(2022)}]{tuzcuoglu2019}
\bibinfo{author}{Tuzcuoglu, K.}, \bibinfo{year}{2022}.
\newblock \bibinfo{title}{Composite likelihood estimation of an autoregressive
  panel ordered probit model with random effects}.
\newblock \bibinfo{journal}{Journal of Business \& Economic Statistics} ,
  \bibinfo{pages}{1--15}\DOIprefix\doi{10.1080/07350015.2022.2044829}.
\bibitem[{Vana and Hornik(2021)}]{pub:djmdr:Vana+Hornik:2021}
\bibinfo{author}{Vana, L.}, \bibinfo{author}{Hornik, K.}, \bibinfo{year}{2021}.
\newblock \bibinfo{title}{Dynamic modeling of corporate credit ratings and
  defaults}.
\newblock \bibinfo{journal}{Statistical Modelling}
  \DOIprefix\doi{10.1177/1471082X211057610}.
\bibitem[{Varin and Czado(2010)}]{Varin09}
\bibinfo{author}{Varin, C.}, \bibinfo{author}{Czado, C.}, \bibinfo{year}{2010}.
\newblock \bibinfo{title}{A mixed autoregressive probit model for ordinal
  longitudinal data}.
\newblock \bibinfo{journal}{Biostatistics} \bibinfo{volume}{11},
  \bibinfo{pages}{127–138}.
\newblock \DOIprefix\doi{10.1093/biostatistics/kxp042}.
\bibitem[{Varin et~al.(2011)Varin, Reid and Firth}]{varin_overview}
\bibinfo{author}{Varin, C.}, \bibinfo{author}{Reid, N.},
  \bibinfo{author}{Firth, D.}, \bibinfo{year}{2011}.
\newblock \bibinfo{title}{An overview of composite likelihood methods}.
\newblock \bibinfo{journal}{Statistica Sinica} \bibinfo{volume}{21},
  \bibinfo{pages}{5--42}.
\newblock \URLprefix \url{http://www.jstor.org/stable/24309261}.
\bibitem[{Varin and Vidoni(2005)}]{Varin2005}
\bibinfo{author}{Varin, C.}, \bibinfo{author}{Vidoni, P.},
  \bibinfo{year}{2005}.
\newblock \bibinfo{title}{A note on composite likelihood inference and model
  selection}.
\newblock \bibinfo{journal}{Biometrika} \bibinfo{volume}{92},
  \bibinfo{pages}{519--528}.
\newblock \DOIprefix\doi{10.1093/biomet/92.3.519}.
\bibitem[{Virolainen(2021)}]{virolainen2021gaussian}
\bibinfo{author}{Virolainen, S.}, \bibinfo{year}{2021}.
\newblock \bibinfo{title}{Gaussian and student's $t$ mixture vector
  autoregressive model with application to the asymmetric effects of monetary
  policy shocks in the euro area}.
\newblock \bibinfo{journal}{arXiv preprint arXiv:2109.13648} .
\bibitem[{Wang and Fan(2010)}]{wang2010ecm}
\bibinfo{author}{Wang, W.L.}, \bibinfo{author}{Fan, T.H.},
  \bibinfo{year}{2010}.
\newblock \bibinfo{title}{Ecm-based maximum likelihood inference for
  multivariate linear mixed models with autoregressive errors}.
\newblock \bibinfo{journal}{Computational Statistics \& Data Analysis}
  \bibinfo{volume}{54}, \bibinfo{pages}{1328--1341}.
\newblock \DOIprefix\doi{10.1016/j.csda.2009.11.021}.
\bibitem[{Zellner and Tiao(1964)}]{zellner1964bayesian}
\bibinfo{author}{Zellner, A.}, \bibinfo{author}{Tiao, G.C.},
  \bibinfo{year}{1964}.
\newblock \bibinfo{title}{Bayesian analysis of the regression model with
  autocorrelated errors}.
\newblock \bibinfo{journal}{Journal of the American Statistical Association}
  \bibinfo{volume}{59}, \bibinfo{pages}{763--778}.
\newblock \DOIprefix\doi{10.1080/01621459.1964.10480726}.

\end{thebibliography}

\end{document}